\documentclass[prb,twocolumn,aps,showpacs,amsmath,amssymb]{revtex4-1}

\usepackage{color}
\usepackage{bm}
\usepackage{graphicx}
\usepackage{braket}

\usepackage[plainpages=false,pdfpagelabels,colorlinks=true,linkcolor=red,urlcolor=blue,citecolor=blue,pdftitle={Title},
pdfauthor={},pdfdisplaydoctitle=true,pdfduplex=DuplexFlipLongEdge]{hyperref}

\def\be{\begin{equation}}
\def\ee{\end{equation}}
\def\ba{\begin{eqnarray}}
\def\ea{\end{eqnarray}}

\begin{document}
\title{Many-body localization in long range \textcolor{black}{hopping} model:  Real space renormalization group study}
\author{Ranjan Modak$^1$ and Tanay Nag$^2$} 
\affiliation{$^1$ SISSA and INFN, via Bonomea 265, 34136 Trieste, Italy}
\affiliation{$^2$ SISSA, via Bonomea 265, 34136 Trieste, Italy}

\begin{abstract}
We develop a  real space renormalization group
(RSRG) scheme by appropriately inserting the  long range
hopping $t\sim r^{-\alpha}$ \textcolor{black}{with nearest neighbour interaction} to study the entanglement entropy and maximum block size for
many-body localization (MBL) transition. We show that for $\alpha<2$ there exists a localization
transition  with
renormalized disorder that  depends logarithmically on the \textcolor{black}{finite size of the system}.
The transition  observed for $\alpha>2$ does not need
a rescaling in disorder strength.
Most importantly, we find that even though the MBL transition for  $\alpha >2$ 
falls  in the same universality class as that of the short-range models, while
 transition for  $\alpha<2$   belongs to a different universality class.
 \textcolor{black}{Due to the intrinsic 
nature of the RSRG flow towards delocalization, MBL phase for $\alpha>2$ might suffer an instability  in the thermodynamic limit 
while the underlying systems support algebraic localization.}
 Moreover, we verify these findings by
inserting microscopic details to the RSRG scheme where we additionally
find a more appropriate rescaling function for disorder strength;
\textcolor{black}{we indeed uncover a power law scaling with a logarithmic correction 
and a distinctly different stretched exponential scaling for $\alpha<2$ and $\alpha>2$, respectively, by analyzing system with 
finite size}. 
\textcolor{black}{This finding further suggests that microscopic RSRG scheme is able to give a hint of instability of the MBL phase 
for $\alpha>2$ even considering systems of finite size.}

\end{abstract}
\maketitle

\section{Introduction}
\label{s1}

Localization-delocalization transition occurring in quantum system separates
 the non-ergodic, reversible phase from the ergodic, irreversible phase of matter \cite{basko.2006,pal.2010}.
The concept of Anderson localization, observed in single particle picture \cite{anderson.1958}, 
can be elevated to many-body localization (MBL) in the interacting
system even in  finite temperature \cite{fleishman.80,altshuler.97}. The intensive investigation of the above phenomenon unfolds many 
unusual response properties \cite{gopalakrishnan.2015, khemani.15}, new nature of quantum entanglement \cite{bardarson.2012, khemani.17,prosen.2008},
and non-trivial phases of matter absent in equilibrium \cite{pekker14}.
{For example, MBL phase violates eigenstate thermalization hypothesis (ETH) \cite{rigol2007relaxation,rigol2008thermalization,santos.2010,rigol1.2009,vidmar_rigol_16}, is characterized by an area law  of entanglement entropy (EE)
and localization length,
while delocalized ergodic phase satisfies volume law for both of them \cite{de2006entanglement,fagotti2011entanglement}.}
On the other hand, in the context of time periodic Floquet system, MBL phase  
can help in exploring the Floquet time crystal \cite{khemani.16,else.16}. Cold atomic systems happen to be a good test 
bed for investigating the MBL transition\cite{serbyn.14,yao.14}.
The experimental search has already began in this field of research to check the theoretical predictions \cite{ehud.2015, choi.16}. 
However, it is important to point out that even though 
analytic perturbative arguments  support the existence
of MBL \cite{serbyn.2013,ros.2015}, very recently the stability of this phase has been  questioned 
in interacting systems with correlated \cite{vznidarivc.2018} and un-correlated disorder \cite{lev.2019}.


It is natural question to ask that whether MBL transition persists  for long range hopping: $t\sim r^{-\alpha}$
as Anderson showed that single particle localization can not occur in the 
presence of long range hopping for $\alpha \le d$ where $d$ is the dimension of the system.   
Most of the numerical attempts in one dimensional system show that MBL can not survive for $\alpha<2$  while 
MBL occurs for $\alpha>2$ \cite{ burin.15,gutman.16,mirlin.18,roy2019self}. Interestingly, perturbative treatment on an effective 
Anderson model can show MBL transition even for $\alpha<1$ \cite{heyl.15}. 
Recently, it has been found that
a different kind of localization namely, algebraic localization, takes place  due to the presence of long-range hopping \cite{deng.2018,deng2018one}, 
which gives rise to  some interesting unique  phenomena such as anomalous transport  \cite{saha2019anomalous,saha2019more} and  
a sub-extensive scaling of EE \cite{modak.19}. {Moreover, in the context of quantum spin chain 
the long range interaction is also investigated in detail leading to a plethora of non-trivial results \cite{pappalardi.18, 
alessio.19a, alessio.19b}.}
These upsurge of theoretical  studies in long range 
model are highly motivated by a series of earlier  experimental investigations \cite{childress2006,dutt2007, Ni.08,miranda.11,korenblit2012,sous2018many,sous2018possible}.


In the context of MBL transition, 
a promising alternative to the existing exact diagonalization(ED) technique is real space renormalization group (RSRG) description
\cite{vosk.15,potter.15,zhang.16, alan.19,anna.19b,potter.17}. The main advantage of using the RSRG technique is that it can overcome
the system size limitation.
{Within this approach, we solve 
a macroscopic version of the underlying model instead of  solving the actual interacting microscopic 
model where the Hilbert space dimension grows exponentially with system size.}
The common principle {employed} in all the RG schemes  is that the collective resonant tunneling processes are energetically favored in the
delocalized phase while localized phase supports the formation isolated islands caused by the suppression of resonant tunneling. 
{Moreover, there has been a recent proposal to incorporate the  microscopic details  in  the RSRG scheme to study  models
with quasi-periodic potential \cite{zhang.18}.} 
Till now all ED results suggests the violation of Harris-Chayes-Chayes-Fisher-Spencer criterion\cite{harris1974effect,chayes.86}, which claims 
the diverging localization length exponent $\nu \geq 2/d$. 
Interestingly, $\nu$ obtained from  RSRG studies satisfies the above criteria.




 Much having explored in the field of  RSRG technique with the short range \textcolor{black}{hopping} model,
our focus here is to extend the RSRG analysis to the interacting long range 
system with hopping as $t\sim r^{-\alpha}$.
The questions that we would like to answer  are  how one can identify and characterize the MBL transition in long range 
\textcolor{black}{hopping} system with $\alpha$ \textcolor{black}{in presence of nearest neighbor interaction}.
Overcoming the system size barrier that {one encounters} in ED, RSRG formalism can  decisively convey that 
the renormalization of disorder strength is essential to observe the true MBL transition in the thermodynamic limit for this kind of system 
with $\alpha<2$. On the contrary, MBL transition for $\alpha>2$ requires no renormalization of disorder strength. Most interestingly,
our analysis with correlation length exponent 
$\nu$ suggests that the universality class of the MBL transition occurring for $\alpha<2$ is different than the usual Anderson type which we observe 
for $\alpha>2$. 
We then strengthen our findings by incorporating the microscopic details in the RSRG scheme where we additionally find  a more 
appropriate renormalization of disorder strength. 
\textcolor{black}{We here find that our numerically obtained scaling functions for the disorder strength are mostly in congruence 
with analytical studies considering long range nature of hopping as well as interaction \cite{gopalakrishnan.19,burin.15}.} 
\textcolor{black}{The microscopic RSRG scheme allows us to get a hint about the instability of the MBL phase for $\alpha>2$
even with the finite size of the system where macroscopic RSRG predicts a MBL transition. It would be an interesting question to study the 
stability of MBL transition for $\alpha>2$ in algebraic localization considering large system size.}

%
 
 We shall now discuss about the organization of the paper. We first introduce the RSRG algorithm for the long range system in 
 Sec.~\ref{s2}. We also present the prescription for the calculation of EE and MBS. We then elaborate on our findings in Sec.~\ref{s3}. We here 
 analyze the behavior of EE and MBS to characterize the MBL transition occurring in finite size system. Next, we briefly 
 discuss the microscopic input to this RSRG scheme and investigate its consequences. \textcolor{black}{We compare microscopic and 
 macroscopic RSRG findings by analyzing the histogram of largest blocks in Sec.~\ref{comparison}.}
 Lastly, in Sec.~\ref{s4}, we conclude. 


\section{RG Scheme}
\label{s2}

We now describe in details the implementation of the RSRG approach,
which we employ here to study the long-range models. 
The main idea is to investigate the structure of  resonance clusters, caused by the destabilization of MBL phase,  using appropriate 
RG rules for our systems. 
Finding  all such generic many-body resonances for a microscopic models  is a challenging problem both analytically and also even numerically. 
Numerical studies  suffer from severe  system  size limitations, because of the exponential growth of Hilbert space dimension with system size $L$. 
Hence,  instead of solving the full resonance structure for any such microscopic Hamiltonian, we first identify small resonant clusters 
starting from two-sites resonance pairs. We  then examine whether groups of these small resonant clusters can collectively 
resonate {or not}. We  apply these techniques iteratively to identify the the structure of resonance clusters in the large scale. 
The RG rules with technical detail, {which is very similar to the one proposed 
by Dumitrescu et al. \cite{potter.17}}, are described below. 
\textcolor{black}{ We at the same time note that RSRG approach can be regarded as a
certain toy model for MBL transition, whether or not the RSRG rules
provide a correct description of MBL transition is still a subject of
debate. The ED provides a full grasp into complex nature of disordered interacting systems which 
might not be fully captured by RSRG schemes.}

\textcolor{black}{We note that the RSRG approach is not limited 
to a certain type of lattice model. However, one can think a corresponding 
microscopic model 
\begin{eqnarray}
 {H}=-\sum _{i,j\neq i} {J_{ij}}(\hat{c}^{\dag}_i\hat{c}^{}_{j}+
 \text{H.c.})+\sum _{i}\mu_i \hat{n}_i + V \hat{n}_i \hat{n}_{i+1}
 \nonumber \\
 \label{eq:micro_ham}
\end{eqnarray}
where $\hat{c}^{\dag}_i$ ($\hat{c}_{i}$) is the fermionic creation (annihilation) operator at site $i$,
$\hat{n}_i=\hat{c}^{\dag}_i\hat{c}_{i}$ is
the number operator.  $\mu_i$ is the 
chemical potential uniformly distributed on the interval $[0,W]$.
$V\ll (J_{ij},W)$ is the strength of nearest neighbour 
interaction. $J_{ij}$ is considered to be the long-range hopping 
between $i$ and $j$-th site leading to the algebraically 
localized SPSs. 
$L$ is the size of the system. It is noteworthy that once 
$J_{ij}=J_{i,i+1}=J$, being restricted to the nearest neighbour only, the
$i$-th SPS is exponentially localized near site $i$
with energy $\epsilon_i 
\simeq \mu_i +{\mathcal O} (J^2/W) $.
In {the  case of  nearest-neighbour interaction only, the diagonal energy mismatch} is  simply given by $\Delta E_{ij} =|\epsilon_i - \epsilon_j|
\sim |\mu_i -\mu_j|$. 
Then the idea of the RSRG scheme 
is to consider the nearest neighbor interaction $V$ as  a  perturbation and 
express the tunneling in terms of SPSs {which are exponentially localized in the above case}.  Hence, the tunneling 
amplitude is having the form $\Gamma_{ij}=V \exp(-|i-j|/x_0)$ with $x_0$
be the localization length\cite{potter.15}.
We would like to mention that for the long-range interacting case \cite{burin.15} with $V_{ij} \hat{n}_i \hat{n}_{j}$, the 
energy mismatch between two sites $i$ and $j$ becomes 
$\Delta E_{ij}\sim |\mu_i -\mu_j + \sum_{k\ne i,j}(V_{ik}-V_{jk}) \hat{n}_k|$. 
In general, the role of 
interactions is to set in  the multi-particle  collective 
resonances.  For  weak  interactions, the system remains in MBL phase
and the local integrals of motion associated
with the weakly dressed single-particle orbitals become 
few-body local integrals of motion \cite{serbyn.2013}.
For  strong enough interactions, MBL phase breaks
down as the clusters become resonantly linked.  
}

\textcolor{black}{Given these details, we shall now implement 
the RSRG scheme with  long-range hopping $J_{ij}$ while the 
interaction $V$ is considered to be nearest neighbour. At the outset, 
we discuss the main philosophy behind our RSRG scheme that we adopt.
For very strong disorder,
the resonantly linked pairs are well separated and their density is 
very small. As the disorder weakened, these  resonant links are
more frequently formed 
 eventually leading towards the disruption of  localization. 
One can note that the process of making  the 
collective resonant links has to be energetically 
favored. These could lead to an avalanche mechanism where 
the system continues to stuck and the size of the resonant 
cluster grows \cite{gopalakrishnan.19,roeck.17}. Therefore, the delocalization stems from the 
these avalanches. Below we shall discuss it with technical detail. }

First, we consider a chain of $L$ sites  and assign each sites with some random number $\lambda_i=[0,W]$ identified
as on-site energy. Next, we  need to initialize the  tunneling matrix elements  $\Gamma_{ij}$, which represent the typical tunneling 
amplitude between $i$ and $j$ sites. As the single-particle wave-function of long-range models are found to be algebraically localized
instead of exponential, we choose   $\Gamma_{ij}=V/|i-j|^{\alpha}$, being our initial values. {Here $V$ can be thought of the nearest 
neighbor interaction strength; we set $V=0.5$ for all our calculations}.
Then, we start our RG procedure by comparing the tunneling matrix elements $\Gamma_{ij}$  between sites $i$ and $j$ with the
energy mismatch $\Delta E_{ij}=|\lambda_i -\lambda_j|$. If $\Gamma_{ij}> \Delta E_{ij}$, we merge those sites 
to build a cluster.   We continue this process iteratively. In each step, $\Gamma$ and $\Delta E$ 
are modified as, $\lambda_{i'}=[\sum_{i} \lambda_{i}^{2}+\sum_{ij}\Gamma^{2}_{ij}]^{1/2}$, 
$\delta_{i'}=\lambda_{i'}/(2^{n_i'}-1)$, and   $\Delta 
E_{i'j'}=\delta_{i'}\delta_{j'}/\min(\lambda_{i'},\lambda_{j'})$, where $i'$ and $j'$ are newly formed clusters and $n_{i'}$ is the number of sites in
cluster $i'$. There is an exception if 
$\lambda_{i'}\geq\delta_{i'}\geq\lambda_{j'}\geq \delta_{j'}$, we then consider $\Delta E_{i'j'}=\min(\delta_{i'}-\lambda_{j'},\delta_{j'})$.
The renormalization rules  of $\Gamma$ during the iterative process are chosen in the following way. 
If two clusters are not modified during a RG step,  the coupling between them is set to zero and  if at least one of the two clusters is modified 
during the RG step, $\Gamma$ is given by,
 
\begin{align}
\Gamma_{i'j'}=\big[\max_{i_1\in \{i\}, i_2 \in \{j\}}\Gamma_{ij}\big]e^{-(n_{i'}+n_{j'}-n_{i_1}-n_{i_2})s_\text{th}/2} \nonumber
\end{align}
where $s_{\rm th}= \ln 2$ is the characteristic entropy per site in the thermal phase.  
  This form is believed to be hold for matrix element of local operators that obey ETH  \cite{potter.17}.
The RG iterative process terminates when no new resonant bond  emerges,  i.e. 
 the cluster structure receives no modifications by further RG steps.

In this paper, we investigate two quantities. 1) Bipartite EE, obtained  by dividing the system into two equal half. 
After the end of a RG procedure for a given initial disorder realization, the  EE is technically 
defined as   $S=\sum_{C}\min(m,n)$. The sum is over all the clusters which span the interval boundary 
and  $m$ , $n$  are number of sites which are separated by the partition of the systems  for such  clusters.
2) Localization length $\xi$, which is defined by the maximum block size (MBS) found  at the end of a RG procedure for a particular 
initial disorder realization. We run our RG procedure for $10^{5}-10^6$ times for  different random realization to obtain average value of EE and
MBS.\cite{potter.17}.
This technique  allows us to simulate system size up to $L\simeq 500$ in contrast to the  ED technique,  which is practically  
impossible {to implement} for any size $L>20$. We analyze  the above  quantities by varying the tunneling exponent $\alpha$ as defined in the 
RG nomenclature. However, considering the underlying physics behind $\alpha$, one can understand that long range (short range) 
corresponds to $\alpha<2$ ($\alpha>2$). \textcolor{black}{However, we note that {the convergence of
our RSRG scheme  is much slower   compared to the RSRG scheme used previously for systems with exponentially localized SPSs\cite{potter.15}.
Hence, our
numerical calculations are limited within the system size 
 $L\sim {\mathcal O}(10^3)$.}}

\textcolor{black}{As passing remarks, we would like to mention that the rules of the RSRG scheme imply
a thermal cluster of size $l_T$ can thermalize spins which are located at
distance $R_{\rm AL}<C ~2^{l_T/a}$ fom the core of the inclusion where $C$ is a constant. The
distance $R_{\rm AL}$ grows exponentially with the size $l_T$ of the thermal inclusion and
the probability of thermalizing the whole system by such a thermal inclusion
quickly reaches $1$ with increasing $l_T$. This means that the strong disorder phase
is, in fact, not stable to inclusion of a sufficiently large thermal region
when the size of the system is thermodynamically large $L\to \infty$.
The uncorrelated nature of the disorder and algebraic structure of single particle states might be responsible 
for the above instability. 
At large $W$, the probability of occurrence of a thermal cluster of size $l_T$ is
very small and the strong disorder phase appears to be localized. However, the
probability grows with system size $L$ eventually leading to thermalization at any
$W$. On the other hand, we would like to emphasize the difference in the RSRG for short-range systems
where a thermal inclusion of size $l_T$ thermalizes only spins residing within 
a fixed region $R_{\rm EL}$ that is related to the 
localization length of the system. Hence this size of $R_{\rm EL}$ would not grow with the system size $L \to \infty$.}

\section{Results} \label{s3}
We here shall describe the behavior of EE and MBS obtained using RSRG scheme 
described  previously.
  We systemically study the critical behavior associated with the transition in Sec.~\ref{s3ss1} and 
Sec.~\ref{s3ss2}.
\subsection{Macroscopic RSRG}
\label{s3ss1}

%



\begin{figure}
\includegraphics[width=0.5\textwidth]{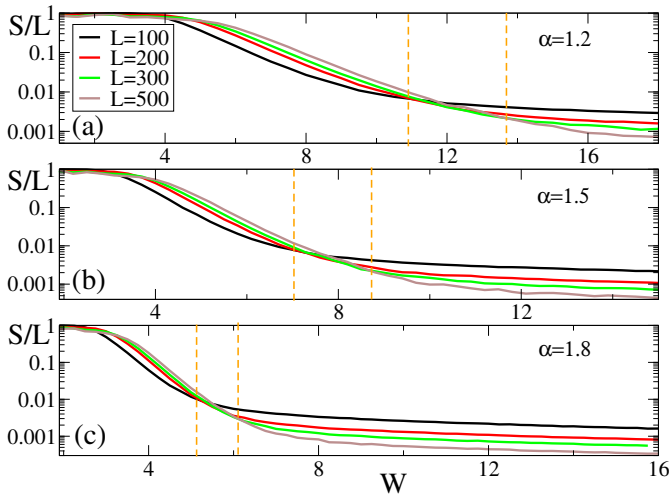}
\caption{(Color online) Plot shows the variation of entanglement entropy (EE) density
$S/L$ as a function of disorder strength $W$ for three tunneling exponents $\alpha=1.2$ in (a), $\alpha=1.5$ in (b)
and $\alpha=1.8$ in (c) with $L=100, ~200,~ 300$ and $500$. The intersection point between 
two consecutive $L$ shifts towards a higher values of $W$ as one increases $W$. The orange dashed line represents the 
disorder window $\Delta W$ within which all the intersection are taking place. One can notice with increasing 
$\alpha$, $\Delta W$ shrinks.  
}
\label{fig1}
\end{figure}

Our aim is to probe the transition with disorder strength $W$ by looking at the
behavior of EE density $S/L$  for different values of $L$. In fig.~\ref{fig1} (a)
with $\alpha=1.2$, we show $S/L$ starts from unity for small 
$W$ and gradually it falls with increasing $W$. $S/L \to 1 (0)$, refers to the fact that the system is in a 
delocalized (localized) phase.
We see that in the small $W$ region, $S/L$ falls more rapidly for smaller $L$ than 
larger $L$ while in the large $W$ region, $S/L$ saturates more quickly for smaller $L$. As a result, we see 
many intersections of $S/L$ between different lengths. A careful analysis suggests that with increasing $L$, the intersection 
between  two  consecutive $L$ shifts towards a higher value of $W$; we refer $W=W_i$ where intersection occurs.
One can demarcate the zone between maximum and minimum value of $W_i$ as $\Delta W$ 
; this is depicted by 
orange dashed line.

Another noticeable observation is that for
$W\gg \textrm{max}\{W_i\}$, $S/L$ does not saturate to a constant value rather their saturation value 
increases with decreasing $L$. This  phase is then no longer a delocalized phase.
For finite size of the system, one can say that there is a delocalization-localization transition if one varies $W$ from 
$W\ll \textrm{min}\{W_i\}$ to $W\gg \textrm{max}\{W_i\}$. 
Therefore, the existence of $W_i$ refers towards a transition but the transition 
points becomes system size dependent.
 We repeat this investigation for $\alpha=1.5$ (see ig.~\ref{fig1} (b)) and $\alpha=1.8$ (see ig.~\ref{fig1} (b))
keeping $\alpha<2$. We observe qualitatively similar feature of the transition but the width of $\Delta W$ shrinks and  $W_i$'s 
shifts towards lower values.
  
We here discuss the uniqueness of this observation.
This is in sharp contrast to the short range lattice models that support exponentially localized single particle states (SPSs)
\cite{luitz.15,zhang.16}. For the above kind of 
model, one can observe a prominent transition point (designated by $W_c$) that does not change with $L$ referring to the fact 
that the  localization-delocalization transition is sharply defined in the finite size system.
It is obviously stable in the thermodynamic limit. On the 
contrary, what we observe here in long range finite size system  for $\alpha<2$
 can better be referred as a crossover. 
{We note that the analogous microscopic long range Hamiltonian supports algebraically localized SPSs \cite{deng.2018}.}
 Precisely, the intersection point $W_i$ is size dependent hence,
 a conventional transition signature between two phases can not be assigned for finite $L$.
 The true existence of the 
  localization-delocalization transition in thermodynamic limit $L \to \infty$ is therefore a subject of 
  investigation  which we shall present below. 
  

\begin{figure}
\includegraphics[width=0.5\textwidth]{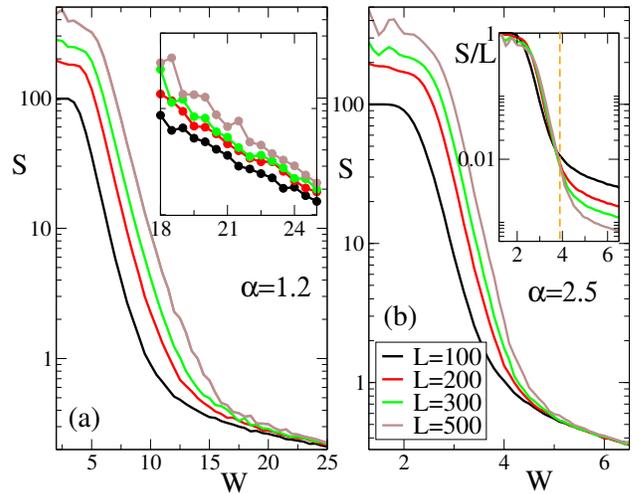}
\caption{(Color online)  The plot depicts the variation of EE $S$ as a function of $W$ for $\alpha=1.2$ in (a)
and $\alpha=2.5$ in (b). (a): One can observe that the system residing in 
delocalized phase ($S\sim L$) meets a transition at $W \simeq 20$ above 
which volume law of EE is no longer satisfied. Inset exhibits a zoomed in figure for $18<W<25$ 
where saturation characteristics of $S$ still shows $L$ dependence. 
(b): It shows a clear transition from delocalized to localized phase 
above $W>W_c=3.95$ where EE satisfies a clear area law. The inset shows the variation of  
$S/L$ with $W$ where  $W_c$ is identified by the dashed orange line.}
\label{fig2}
\end{figure}

Having discussed the situation with $\alpha<2$, we shall now focus 
on $\alpha>2$ sector. 
{In this case, the system is expected to show similar behavior  as compared to the  short range models\cite{mirlin.18}.} 
We compare the behavior of $S$ between $\alpha=1.2$ (see Fig.~\ref{fig2}(a))
and $\alpha=2.5$ (see Fig.~\ref{fig2}(b)). 
For $\alpha=1.2$ case, $S$ for different
$L$ does not show any coincidence for  larger values of $W$; although, EE
shows a tendency towards saturation where saturation value increases with increasing $L$
(see the inset for Fig.~\ref{fig2}(a) where 
a zoomed  version of $S$ is plotted for $18<W<25$).  A clear distinction is seen 
in $\alpha=2.5$ where $S$ for all $L$ coincides with each other for $W > W_c$.
 Once again We emphasize that  for $\alpha=2.5$, the inset of Fig. ~\ref{fig2}(b)  depicts a sharp transition point  $W_c$ for  all values of $L$ similar to one observes for short-range models. While 
 comparing with  Fig.~ \ref{fig1}(a), it is clear that crossover occurs as $W_i$ becomes a function of $L$.

 One can hence infer that the nature of the phase transition 
 even in finite size system changes from $\alpha<2$ to $\alpha>2$ as far as the saturation characteristics of $S$ 
 is concerned. The nature of phase transition for $\alpha>2$ refers to the fact that 
 EE obeys area law for $W>W_c$. On the other hand, for $\alpha<2$, EE apparently follows a sub-extensive scaling violating the 
 area law \cite{modak.19}; this is a very unconventional outcome for a localized phase. 
 Therefore, the natural question comes whether this saturation behavior is an artifact of the 
 crossover.

\begin{figure}
\includegraphics[width=0.22\textwidth]{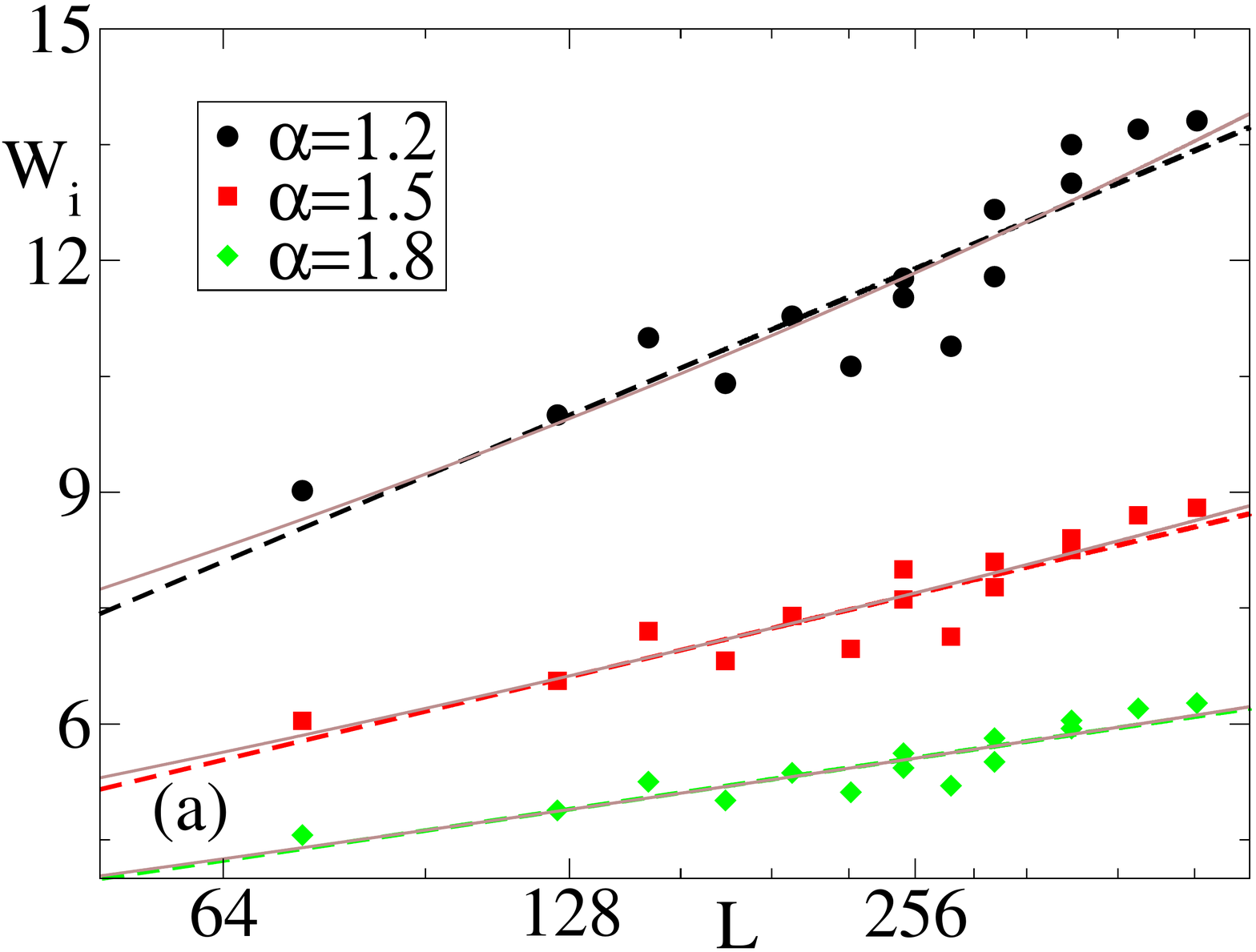}
\includegraphics[width=0.22\textwidth]{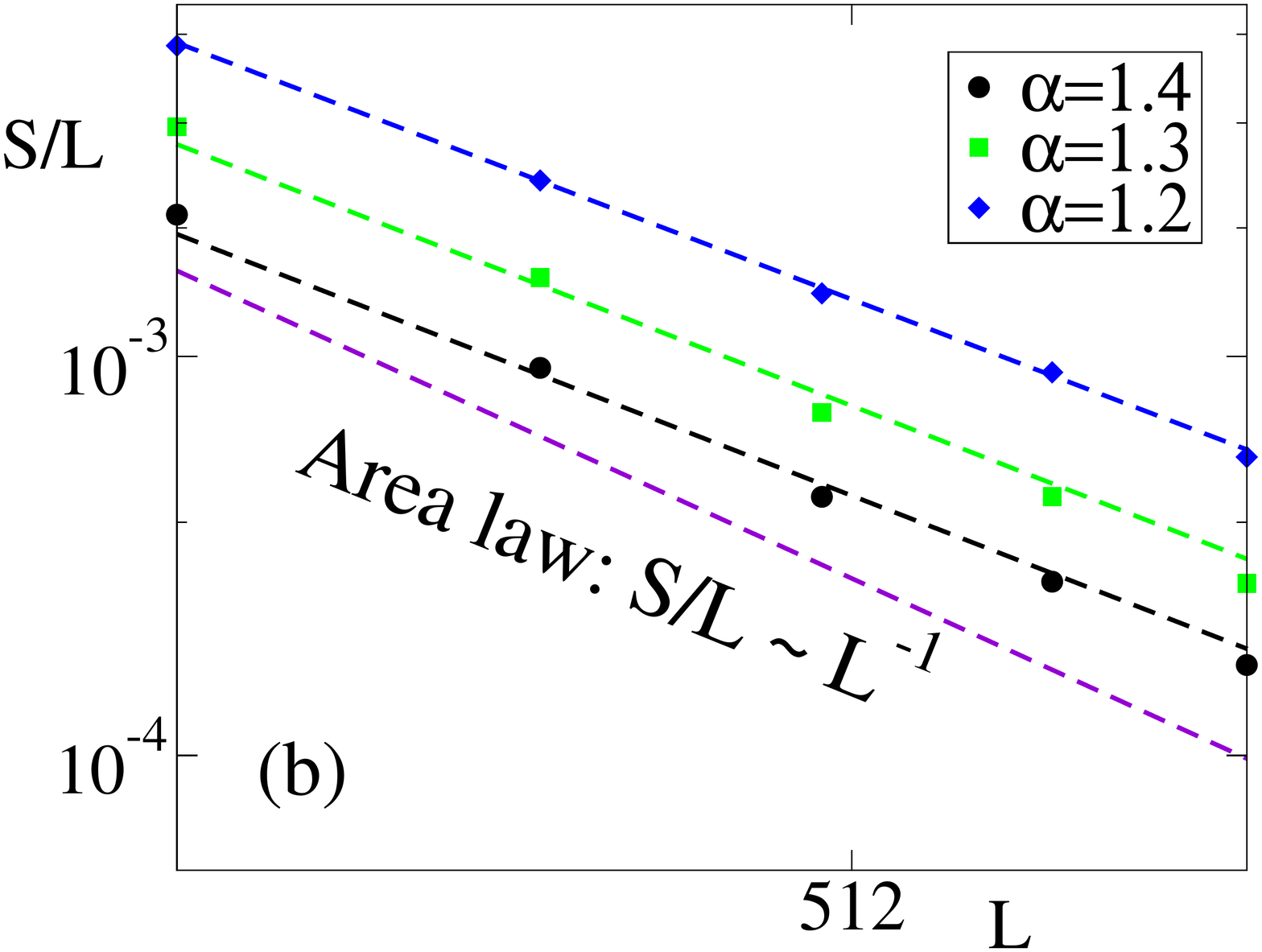}
\caption{(Color online)
(a): We show the scaling of intersection points $W_i$ (solid back, red and green points), obtained 
from crossing between different $L$, with $L$ (in log scale) for 
 $\alpha=1,2,~1.5$ and $1.8$. The  dashed best fit straight lines 
 show that $W_i \sim \gamma \ln L$ where $\gamma$ depends on  $\alpha$.
 \textcolor{black}{The brown solid lines represent the $L^\eta\ln L$ fitting of $W_i$
 where we find $\eta \ll (2-\alpha)$; $\eta=0.11,0.08,0.05$ for $\alpha=1.2,1.5,1.8$.}
(b): We show the scaling of EE density, 
depicted by solid black, green and  blue points, with $L$ (in log scale) with $W=20$ for $\alpha=1,2,~1.3$ and $1.4$.
 The dashed violet straight lines confirms the area law in the localized phase with $W^*=3.0$.
 }
\label{fig3}
\end{figure}

Having compared $S$ between  $\alpha<2$ and $\alpha>2$, we turn our 
focus to investigate about the intersection point more extensively.
One can notice that $S/L$ for a given $L$ intersects with all the other $L$ (denoted by $L'$) in many different 
positions as denoted by $W_i(L,L')$.
The prescription that we are following is $W_i(L,L') \to W_i((L+L')/2)=W_i(L)$; as a result, we get a large set of 
data points which helps in describing 
the feature of $W_i$ with $L$ more precisely. \textcolor{black}{ {
We have also checked the robustness of our results by  adopting different methods for identifying $W_i$ e.g., geometric mean $W_i(L,L')\to W_i(\sqrt{LL'})=W_i(L)$.}
These different  scheme lead to the same scaling function of $W_i$ as reported below.
One can, on the other hand, choose $W_i(L, L+100)\to W_i(L+50)=W_i(L)$ as in our analysis we have $L=100,200,300,400,500$.}
Figure~\ref{fig3}(a) clearly suggests the intersecting points $W_i$ logarithmically scales with $L$: 
$W_i\sim \gamma \ln L$ for $\alpha<2$. However, 
the prefactor $\gamma$ depends on $\alpha$.
\textcolor{black}{However, we also examine the $L^\eta \ln L$ scaling of $W_i$; the best fit value of $\eta$ appears to  be 
{dependent on $\alpha$, but 
 $\eta \ll (2-\alpha)$. Note that $\eta=2-\alpha$ has been predicted earlier in long-range interacting models\cite{gopalakrishnan.19,burin.15}.}
As far as the merit of the scaling is concerned, we check that our results remain unaltered within numerical accuracy 
considering both of the scaling function. Furthermore,  
unless $L$ is chosen to be thermodynamically large,  one really  can not  distinguish 
between these two scaling function.  The power law scaling $L^{\eta}$ might prevail for large $L$. 
 }
Both of these scaling apparently prohibits the transition to happen in the thermodynamic limit: $L\to \infty$, 
$W_i \to \infty$. On the other hand, $\gamma$
approaches zero as one approaches $\alpha=2$; this conveys the fact that there exists a sharp transition 
point $W_c$ which is independent of $L$. Hence, the transition obtained for $\alpha>2$ is  
stable at least in finite system of size  $L\sim O(10^2)$. \textcolor{black}{In the thermodynamic limit $L\to \infty$, the stability of 
MBL phase might be obstructed as the length of the influence region $R_{\rm AL}$ grows with the length of the thermal block $l_T$ (see Sec.~\ref{s2}, for detail 
discussion).  }

\begin{figure}
\includegraphics[width=0.5\textwidth]{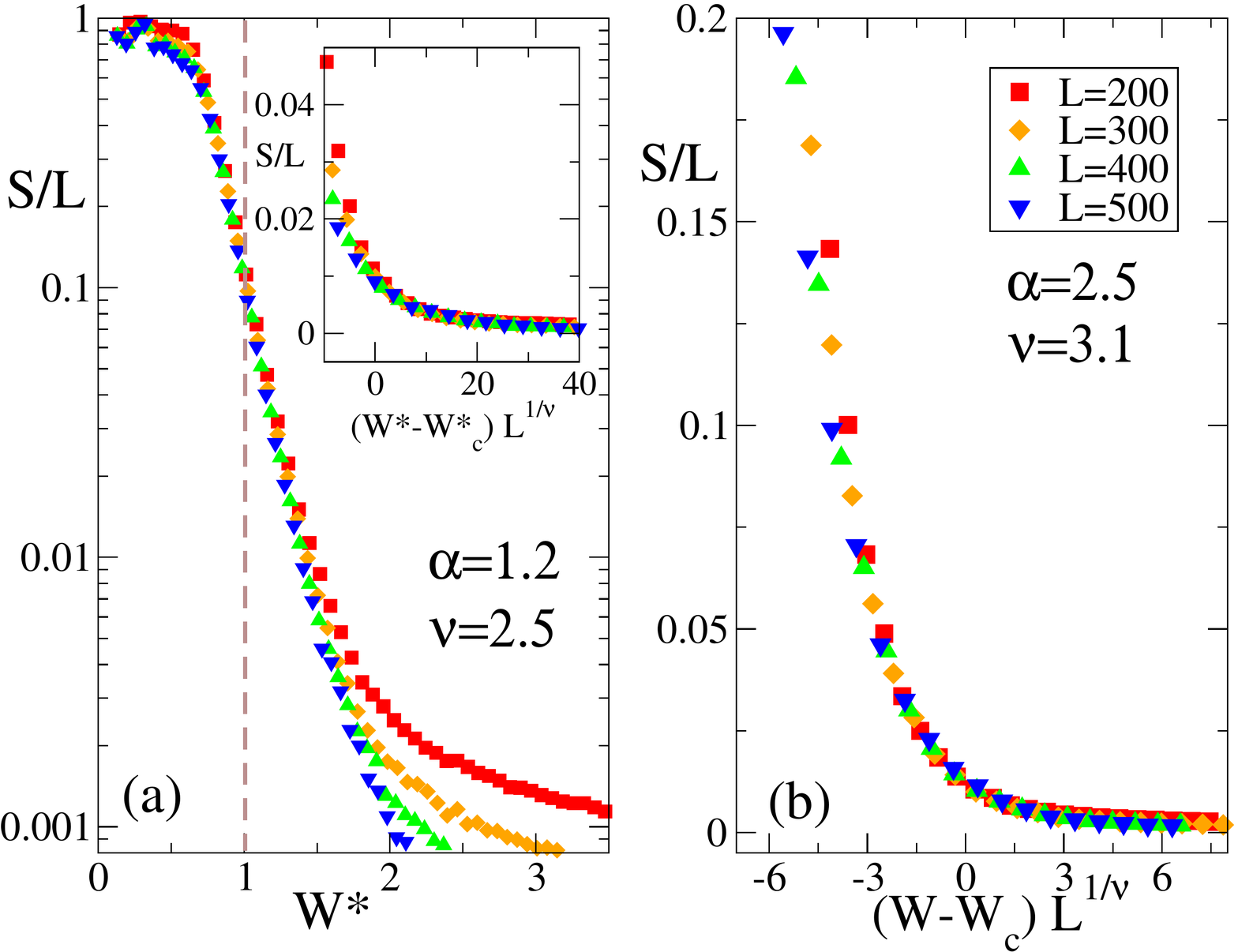}
\caption{(Color online)   We here show the finite size scaling analysis for 
$\alpha=1.2$ in (a) and $\alpha=2.5$ in (b). (a): In order to identify 
a sharp transition point, we plot $S/L$ as a function of rescaled disorder $W^*=W/\ln L$.
The coincidence of $S/L$ for different $L$ is observed $W^*<W^*_c=0.99$ after 
which $S/L$ tends to deviate from each other and eventually saturates for $W^*\gg W^*_c$; this saturation value increases as 
$L$ decreases. Inset shows a clear data collapse around $0$ (maximally for $W^*> W^*_c$) when 
$S/L$ is plotted as a function of $(W^*-W^*_c)L^{1/\nu}$ with $\nu=2.5$.
(b): We repeat the inset of (a) in (b) for $\alpha=2.5$ considering the fact that 
$W_c=3.95$. We find a perfect data collapse around $0$ when 
$S/L$ is plotted as a function of  $(W-W_c)L^{1/\nu}$ with $\nu=3.1$.
These clearly suggests that the characteristics of transitions undergoing for 
for $\alpha<2$ and $\alpha>2$ are qualitatively different. }
\label{fig4}
\end{figure}

We are now in a position to investigate the crossover phenomena. Instead of considering the bare 
$W$, we can continue our analysis with the renormalized $W$ namely, $W^*$  
 according to the numerically predicted scaling
\be
W^* =\frac{W}{\gamma L^{\eta}\ln L}.
\label{eq:w_rescale}
\ee
The motivation behind this renormalization is  to identify the proper transition point $W^*_c$  for 
a thermodynamic system.
Figure~\ref{fig4}(a) depicts the 
variation of $S/L$ as a function of rescaled disorder $W^*$ with $\alpha=1.2$. There one can clearly  notice the 
existence of a critical point $W^*_c$ separating the delocalized phase from the localized phase.
We shall extensively describe below this observation with plausible argument.

We now probe the saturation scaling of EE with $L$ for a large but fixed value of $W$ and $W^*$ simultaneously.
Figure~\ref{fig3}(b) apparently suggests that for $\alpha<2$,
{ $S/L$  scales as $L^{-\eta}$ with $\eta\simeq 0.9$  in the large $W>\textrm{max}\{W_i\}$ limit
as depicted by the solid point symbols. Even though, this observation is in congruence with the non-interacting case of the microscopic 
model \cite{modak.19},
the scaling exponent however  remains almost independent of the choice of $\alpha$ unlike the non-interacting case. 
This might be due to the mixing of the Hilbert space degrees of freedom for an interacting system. 
Moreover, adiabatic continuity demands that in the weakly interacting case, EE should also obey the sub-extensive law in the localized phase.
Our macroscopic RG scheme might not be sufficient for studying this law which is deeply governed by the microscopic nature of the model.}
However, this outcome goes against the usual notion of localized phase
in the context of MBL transition. We hence scrutinize our observation by considering the proper renormalized $W^* \gg W^*_c$.
This restricts us to stay well inside the localized phase rather than in the vicinity of the crossover region. We there observe an 
absolute area law of EE in the finite size system that is depicted by violet dashed line in Fig.~\ref{fig3}(b).
{Therefore, irrespective of the microscopic nature the 
renormalization of $W$ again becomes relevant to observe the accurate behavior
associated with a MBL phase (we discuss this at length in Sec.~\ref{s3ss2}).}
\textcolor{black}{We would like to emphasize that the deviation from the area law for $\alpha<2$ is not an artifact 
of the RSRG scheme with nearest neighbour interaction. It  rather might be intrinsically caused by the 
algebraic nature of the SPSs \cite{modak.19}. }
\textcolor{black}{ As discussing the area law of MBL phase, 
it is also important to highlight a major difference between the mechanisms of thermalization for long-range models
in comparison to the short-range models. In short range systems,
  a thermal inclusion  thermalizes only spins in a fixed range depending  on the localization length. 
Contrastingly, for long-range systems  a such thermal inclusion will thermalize the whole systems.  It means that
delocalization is much more  favorable for long-range models. }


Turning to the Fig.~\ref{fig4}(a), 
the visual inspection shows that the rescaling of disorder strength leads to an approximate coincidence of all 
rescaled curve up to a certain point $W^*_c=0.99$; $W^*>W^*_c$, the coincidence is lost and they
start deviating from each other
and $S/L$ saturates to a higher value as $L$ increases. 
Therefore, one can obtain a sharp transition point $W^*_c=0.99$. 
Having obtained $W^*_c$, one can check the finite size exponent $\nu$, following  
the data collapse technique near the transition point for $\alpha=1.2$.
Our focus would be obtain a proper collapse in the right side of $W^*_c$ i.e., $W^*>W^*_c$ as the region
 $W^* < W^*_c$ is less of our interest. The functional form that we keep 
in our mind is $S/L=f((W^*-W^*_c)L^{1/\nu})$ near the transition point. 
We show in the inset of Fig.~\ref{fig4}(a) that $(W-W^*_c)L^{1/\nu}$ plots of $S/L$ for different $L$
coincide (maximally for $W^*\ge W^*_c$) with each other near $0$ with $\nu=2.5\pm 0.31$. To show the robustness of this exponent, we consider different 
interaction strength $V$ and $\alpha<2$. We find remarkably that critical exponent obtained for various settings are well inside the 
error bar.

On the other hand, we perform a data collapse for $\alpha=2.5$ in Fig.~\ref{fig4}(b) keeping the same 
mathematical form in our mind  near the transition point: $S/L=f((W-W_c)L^{1/\nu})$. We note here
that the renormalization of $W$ is no longer required as the a sharp transition point is obtained 
from  bare $W$  unlike the case for 
$\alpha=1.2$. Moreover, the localized phase obey area law for $W\gg W_c$.
The interesting observation 
is that with $\nu=3.1\pm 0.25$, one can obtain a very nice data collapse 
around $W=W_c$ in both the sides.


Extraction of these exponents conveys a lot of physical message 
about the transition for $\alpha<2$ and $\alpha>2$. The transition observed for $\alpha=2.5$
is qualitatively different from the one observed 
for $\alpha=1.2$ as far as the critical exponents are concerned. However, the localized phases obtained for 
both sides of $\alpha=2$ bear the signature of area law. The nature of data collapse we observe in Fig.~\ref{fig4}(a) with 
$\alpha=1.2$ allows us to 
convey the message that there might be two different critical exponents present in left and right side of 
$W^*_c$ \cite{mirlin.18}. 
By invoking the concept of correlation length $\varepsilon$ in the Hilbert
space of the problem
near the transition point, we can write down the following scaling
relation
\ba
\varepsilon(W*) &\sim&  (W^*-W^*_c)^{-\nu_1}, ~~~\textrm{for} ~~~ W^* > W^*_c,\nonumber\\
\varepsilon(W*)&\sim & (W^*_c-W^*)^{-\nu_2},~~~\textrm{for} ~~~W^* < W^*_c.
\label{eq:RG_xi}
\ea
Here we consider $\nu_{1,2}$ being the correlation length exponent, when 
 $W^*_c$ is approached from above i.e., localized  (below i.e., delocalized)
phase. This is clearly not the case for $\alpha>2$ as shown in Fig.~\ref{fig4}(b) where a single exponent $\nu=3.1$ can decisively 
collapse all $S/L$ curves for different $L$.

The existence of 
two different $\nu$ in two sides of transition point might be related to the absence of proper length scale namely, localization length inside the 
system. Additionally, near the critical point in disordered system, there exists Griffiths phase
\cite{griffiths.69}; this idea is also extensively explored in the context of MBL transition \cite{vosk.15}.
One also needs to consider the effect of Griffiths phase in describing these exponents. 
However, what we would like to 
emphasize more is that for $\alpha<2$,  system essentially being long range {(we reiterate that 
SPSs of a microscopic long range  Hamiltonian are algebraically localized)}
\cite{deng.2018},
the critical behavior associated with the MBL transition suggests that it belongs to a different universality class as 
compared to the MBL transition occurring 
for $\alpha>2$. On the other hand, the MBL transition happening for $\alpha>2$ belongs to the Anderson type universality class 
for short range system where SPSs are exponentially localized \cite{luitz.15}. We note that $\nu$ satisfies Harris criteria \cite{harris1974effect}
for the MBL transition  in both the sides of $\alpha=2$. It is worth mentioning that 
the change in the universality class is also visited in the field of quantum spin chain 
where the range of spin-spin interaction is tuned \cite{dutta.01}. It is indeed a strength of the RG analysis that even
without considering a microscopic Hamiltonian, it can signal 
the change in the universality class while range of tunneling matrix element $\Gamma_{ij}$ is varied.


\begin{figure}
\includegraphics[width=0.23\textwidth]{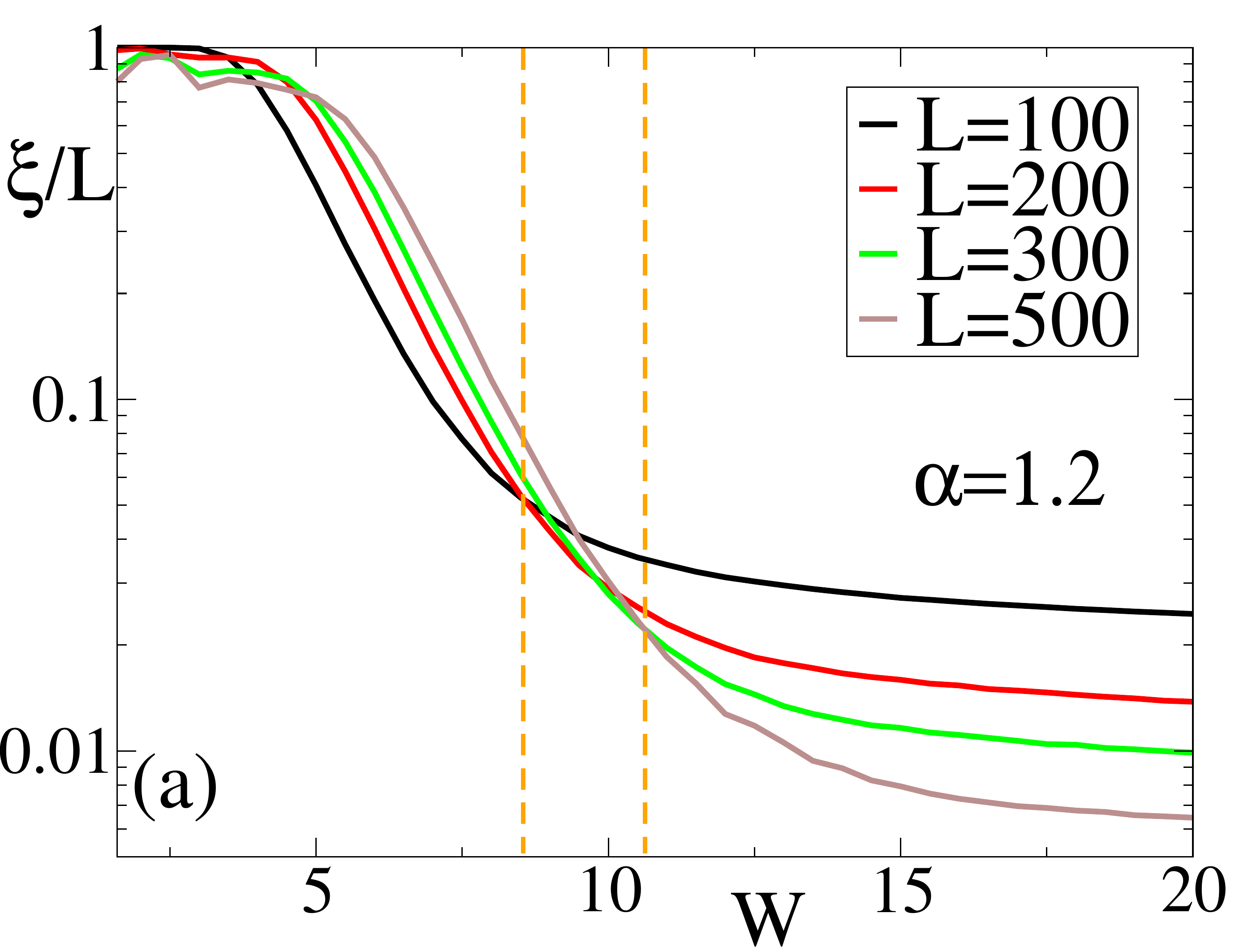}
\includegraphics[width=0.23\textwidth]{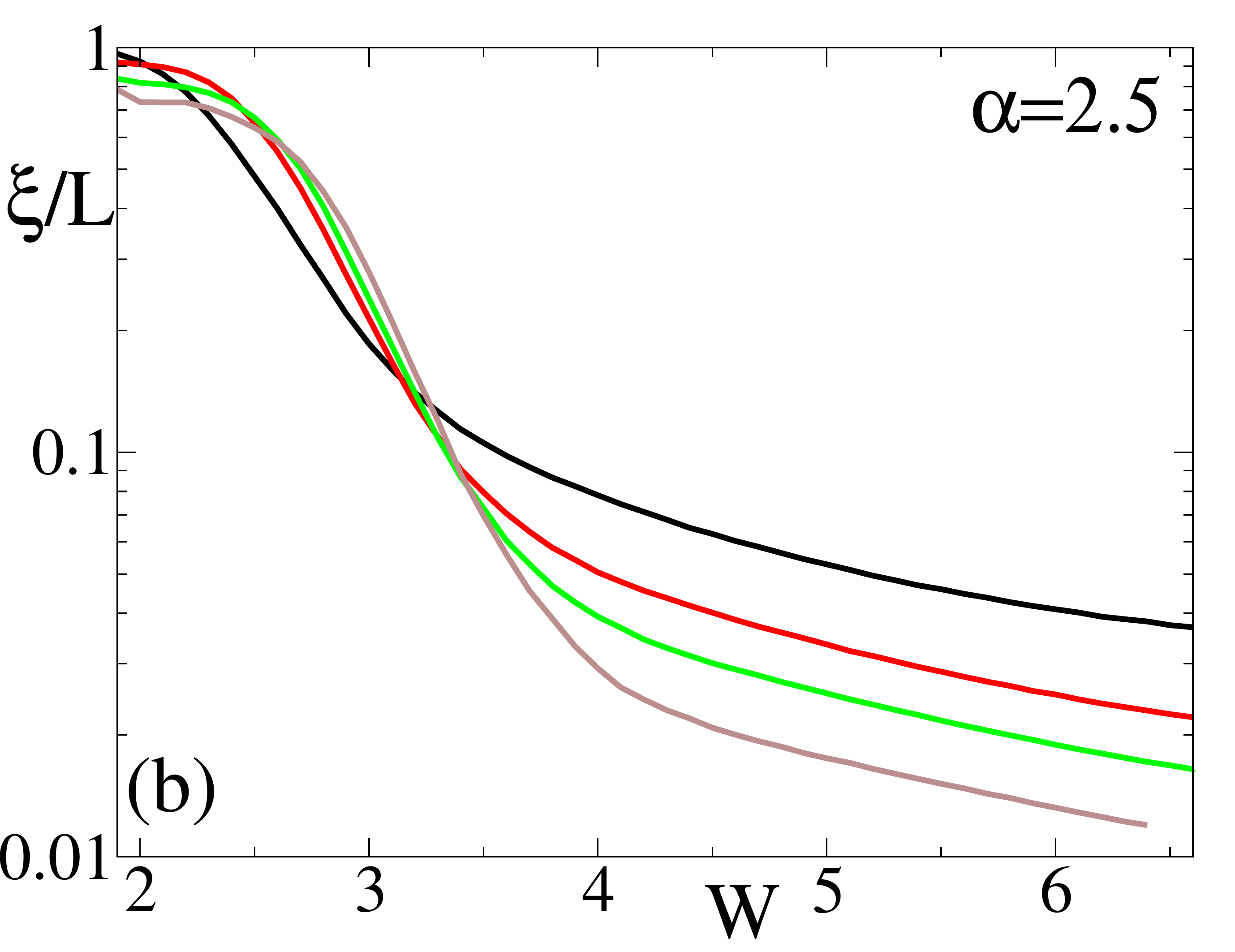}\\
\includegraphics[width=0.23\textwidth]{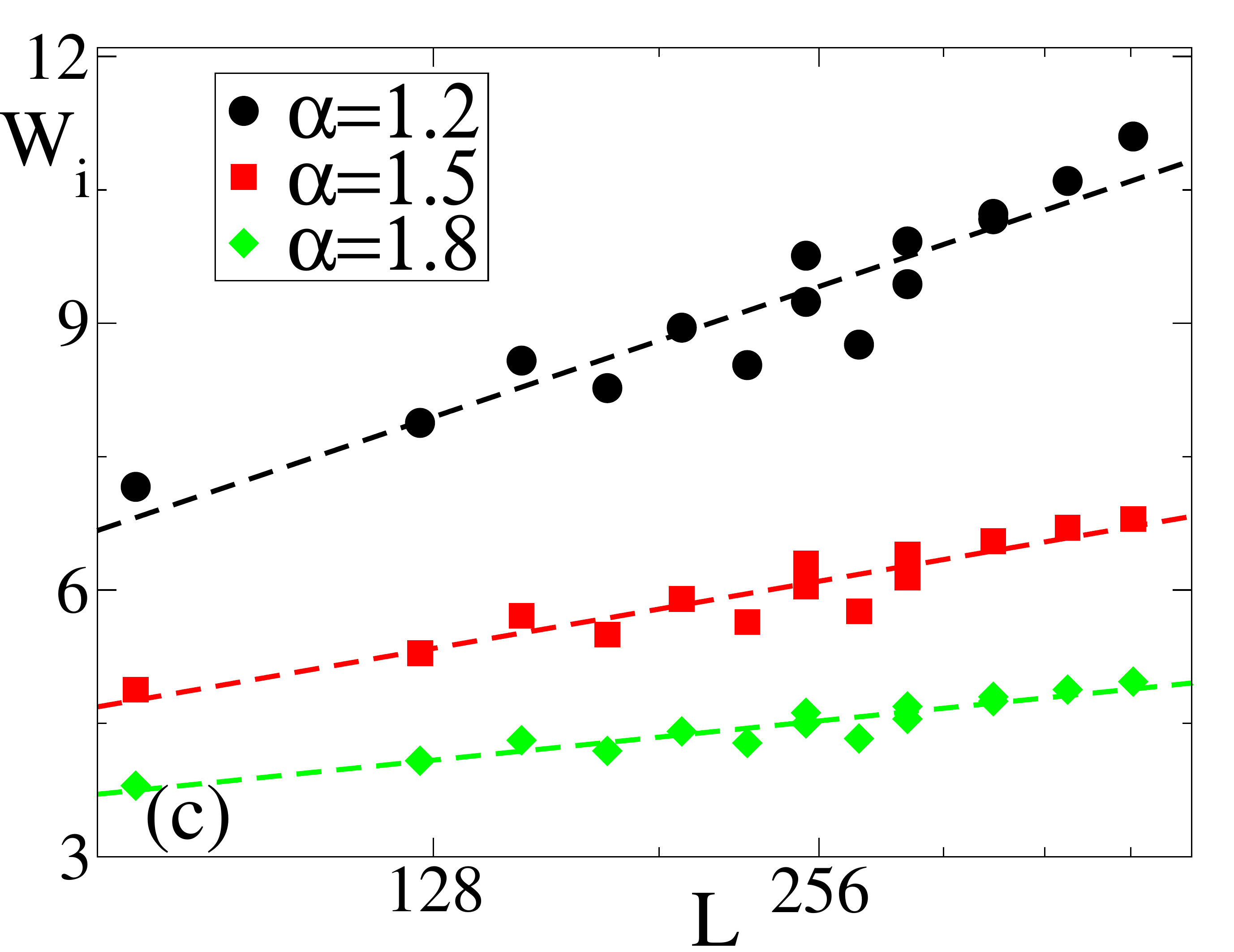}
\includegraphics[width=0.23\textwidth]{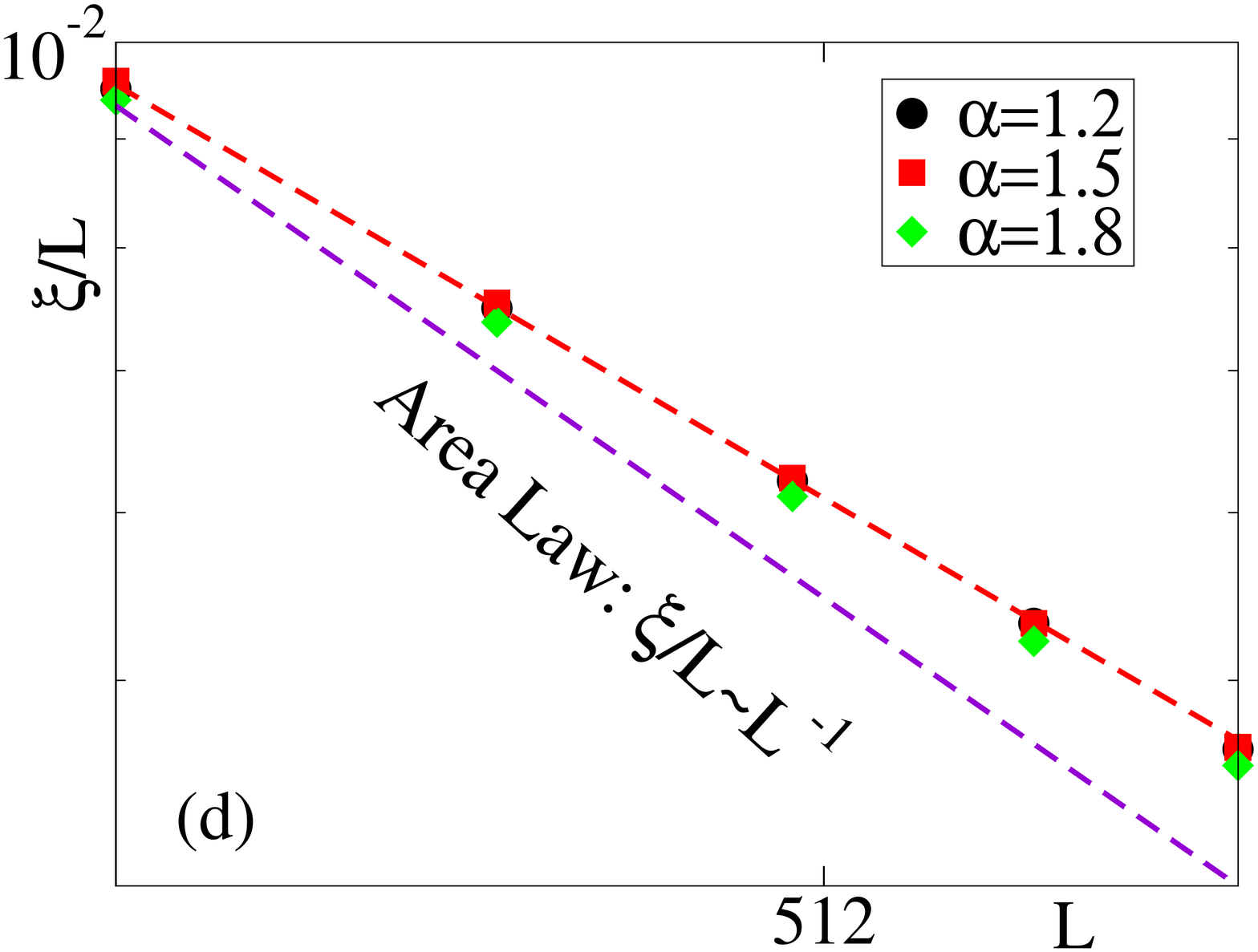}
\caption{(Color online) We here present the findings for maximum block size (MBS). (a):normalized MBS $\xi/L$ shows a transition from 
high value (delocalized phase) to low value (localized phase),
although, the intersection points shifts to higher values of $W$ as 
one increases system size from $L=100$ to $500$. Here $\alpha=1.2$. 
(b): We repeat (a) for $\alpha=2.5$. Remarkably we find that there 
exists a sharp transition point $W_c=3.3$ irrespective of the system size $L$.
(c): We show the intersecting $W_i$ follows similar logarithmic
scaling with $L$ (depicted by the best fit dashed lines) as obtained from EE: $W_i=\gamma \ln L$. $\gamma$
reduces with increasing $\alpha<2$. The solid red, black and green are numerical data points.  
(d): We show a scaling of $\xi/L$ with $L$ in a semi-log scale with $W=20.0$ as depicted by solid red, black and green points. 
The area law is guaranteed in the localized phase for $W^*=3.0$ as shown by the violet dashed line.} 
\label{fig5}
\end{figure}

We shall now investigate the behavior of normalized 
maximum block size (MBS) $\xi/L$. 
As stated above, 
MBS acquires the value of $1$ in the delocalized phase while 
in the localized phase $\xi/L \to 0$.
Let us begin by analyzing the Fig.~\ref{fig5}(a) and Fig.~\ref{fig5}(b)
where $\xi/L$ is shown for different $L$ with $\alpha=1.2$ and 
$\alpha=2.5$, respectively. Findings suggest that delocalization 
($\xi/L \sim 1$) to localization ($\xi/L \ll 1$)
crossover is undergoing for all values of $L$ if we increase $W$
sufficiently $W>\textrm{max}\{W_i\}$. The intersection window $\Delta W$
appears a bit earlier than the one observed in EE for $\alpha=1.2$.
Figure~\ref{fig5}(c) clearly indicates that $W_i$ logarithmically 
scales with $L$.
The crossover in the finite size system would corresponds to a proper transition phenomena if 
we renormalize $W$ following the same scaling formula (\ref{eq:w_rescale}); 
similar to the case of EE, here also one can define a sharp transition point and data collapse. 
On the other hand, $\xi/L$ shows a 
clear transition point at $W_c=3.3$ for $\alpha=2.5$ without any renormalization of disorder strength.
Therefore, the qualitative differences between these two transitions occurring for $\alpha<2$ (i.e., long range limit) 
and $\alpha>2$ (i.e., short range limit) are  also visible from MBS analysis.  
Lastly,
Fig.~\ref{fig5}(d) suggests that proper renormalization of $W$ can guarantee the area law (depicted by violet dashed line)
in the localized phase for $\alpha<2$; the deviation from area law is an artifact of the finite size crossover phenomena. 


We shall now make resort to an analytical formulation where one can 
qualitatively understand the crossover phenomena in the finite size system \cite{burin.15}. 
Let us begin by considering a $d$-dimensional 
hypercube disordered long range model with $N$ (equivalent to $L$) interacting spin-$1/2$ particle with 
spatial density $n$. Here two spins are separated by $R$.
Now the notion of the resonant pair comes in the picture when the system resides in a delocalized ergodic phase i.e., 
spins at different sites can club together and behave as a collective spin. In this phase, the probability to 
form a resonant pair is  $P(R) \sim U_0/(W R^\alpha)$ where $U_0$ the energy scale spin-spin interaction and $W$ is the 
disorder strength.
The density of resonant pair of size $R$ is then given by 
$n_p(R) \sim n R^d P(R) \sim n W^{-1} R^{-\alpha +d} $. Hence the total number of spin within a volume of $R^d$ becomes 
$\tilde{N}(R)\sim n W^{-1} R^{-\alpha +2d}$.
The effective average distance is 
given by $n_p^{-1/d}$. The effective interaction within this average distance then takes the form 
\begin{align}
V(R) \sim \frac{1}{\bigl(n^{-1/d}_p\bigr)^\alpha} \sim W^{-\frac{\alpha}{d}} R^{\frac{\alpha(-\alpha+d)}{d}}
\sim W^{-\frac{\alpha}{d}} \big[\frac{1}{R^d}\big]^{\frac{\alpha(\alpha-d)}{d^2} }.
\label{eq:ana_scaling1}
\end{align}
On the other hand, the characteristic energy of such pair given by 
\be
E(R) \sim R^{-\alpha}\sim \frac{1}{\bigl( R^d \bigr)^{\frac{\alpha}{d}}}.
\label{eq:ana_scaling2}
\ee

Therefore, combining these two energy scales in Eq.~ (\ref{eq:ana_scaling1}) and Eq.~(\ref{eq:ana_scaling2}), 
one can infer that the resonance can only proliferate if effective interaction exceeds the 
characteristic energy. The condition we obtain then 
\be
\frac{\alpha(\alpha-d)}{d^2} < \frac{\alpha}{d}.
\ee
This can be simplified as $\alpha<2d$. One can thus argue that delocalization can take place
for sufficiently large  system $L\gg R$ when $\alpha< 2d$. In the above argument we concentrate only on the 
exponent associated with $R$ and subside the influence of disorder strength $W$. 
Therefore, the limit of large $L$ limit 
requires proper scaling of $W$ with the system size. What we mean by that is the following:
for a given disorder strength $W$, tendency towards 
delocalization increases 
with $L$ and, equivalently, for a given system size $L$,  tendency towards localization increases with increasing $W$. 
Therefore, critical length $L_c(W)$ or critical disorder $W_c(L)$ both can exist. 
The above line of argument further suggests that  if the 
system size becomes comparable with the size of the resonant pair $L \simeq R$ and $\tilde{N}(L)\sim 1$: one can obtain 
$W_c (L) \sim L^{(2d-\alpha)/\alpha}$
and similarly, $L_c(W)\sim W^{\alpha/(2d-\alpha)}$. Hence, interestingly, for finite $L$ true localization transition  
happens to be a crossover. 
Moreover, for $\alpha<2d$, $\tilde{N}(L)$ ranges 
from very small value to large value as $L$ varies from very small value to large value referring to the fact that 
many-body delocalization transition is taking place.
If thermodynamics limit is taken by considering $W$ and $L$ both simultaneously to infinity keeping 
$W/W_c(L)$ fixed, one obtains localized phase for $W>W_c(L)$ and delocalized phase for $W<W_c(L)$. One can connect it to 
phase diagram obtained in $W-L$ plain as represented in {Ref.\cite{mirlin.18}}.

%


 Now the interesting question is how much it is true that  
$W_c$ follows an algebraic scaling with $L$. The resonances occurring inside the system are not of very simple type rather the 
emerging network of the many-body states coupled by these resonances has a treelike structure. To be precise, resonant 
structure in the many-body Hilbert space can be viewed as a random regular graph \cite{gutman.16, mirlin.16,mirlin.18}.  Moreover, the 
 resonances can be identified distinctly from those encountered on the previous step
 resulting in an emergence of spectral diffusion factor 
 \cite{gornyi.17}. Under these circumstances, lattice with 
 connectivity $K\gg 1$, the critical value of disorder enhanced by a factor of $\ln K$. Generally, for a lattice of size $L$, 
 $K=f(L)$. Hence, $W_c$ should contain a logarithmic and an algebraic factor dependent on $L$. However, 
 in this present case, we find $W_c$ scales as logarithmically. This could be an artifact of finite size limitation. As we know, 
 in the large $L$ limit $\ln L$ can be suppressed by the algebraic factor while in the small $L$ limit, $\ln L$ would 
 be predominant over the algebraic factor. \textcolor{black}{ We additionally note that in our case of nearest neighbour short 
 range interaction might not be able to capture the accurate scaling of critical disorder with finite size of the system; furthermore,
 long range interaction might enhance localization and lower the critical disorder as compared to the short range 
 interaction \cite{roy2019self,gopalakrishnan.19}. The MBL phase is found to be destabilized once the long range interaction 
 decays slower than exponentially \cite{roeck.17}. 
 \textcolor{black}{We would like to reiterate that in the thermodynamic limit $L\to \infty$,  MBL phase for $\alpha>2$,
 obtained following macroscopic RSRG scheme, might suffer from instability. The algebraic nature of the coupling  i.e., algebraic localization 
 can cause this instability. }
 However, from numerical ED study it is not very conclusive that 
 what is the effect of long range interaction on MBL transition \cite{nag.19}.}


\begin{figure}
\includegraphics[width=0.23\textwidth]{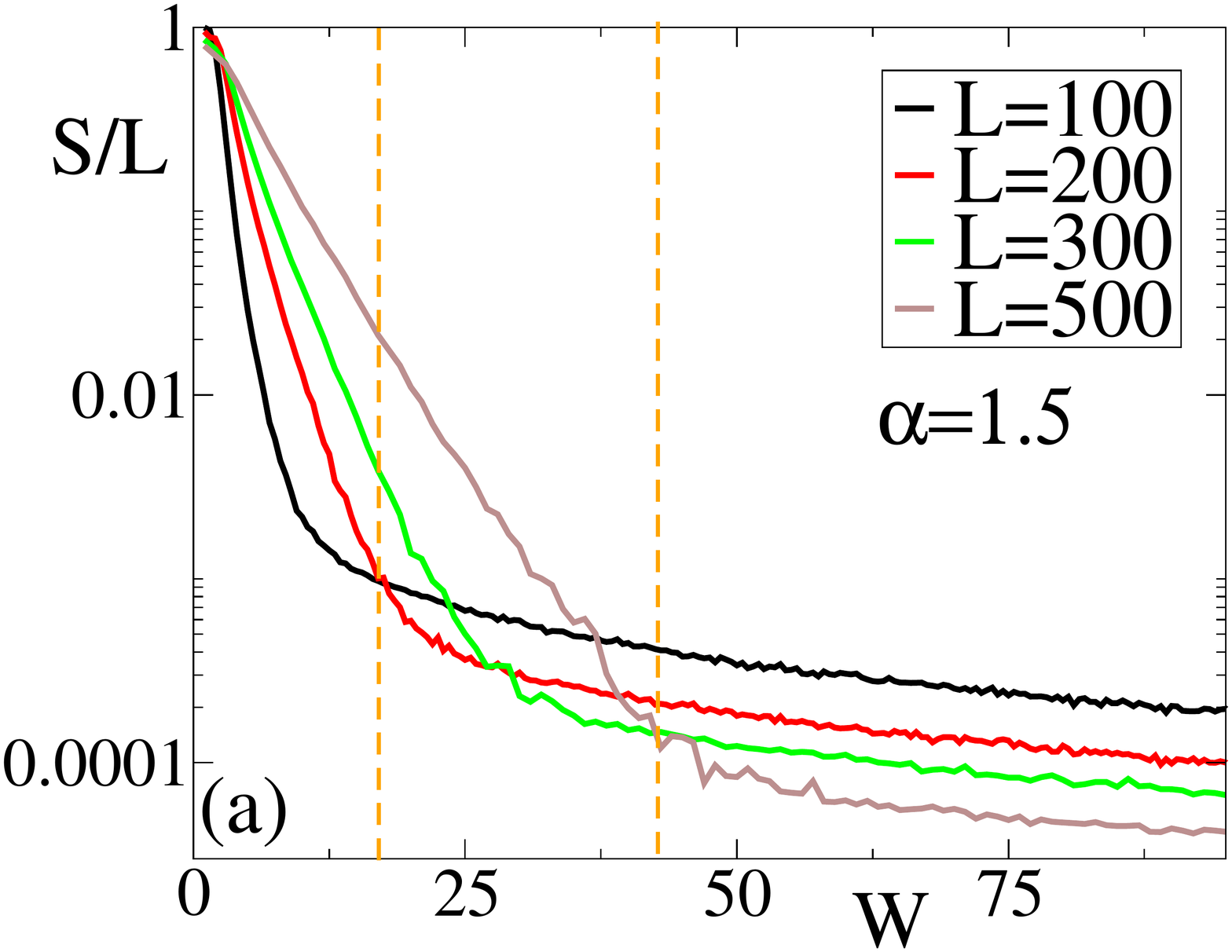}
\includegraphics[width=0.23\textwidth]{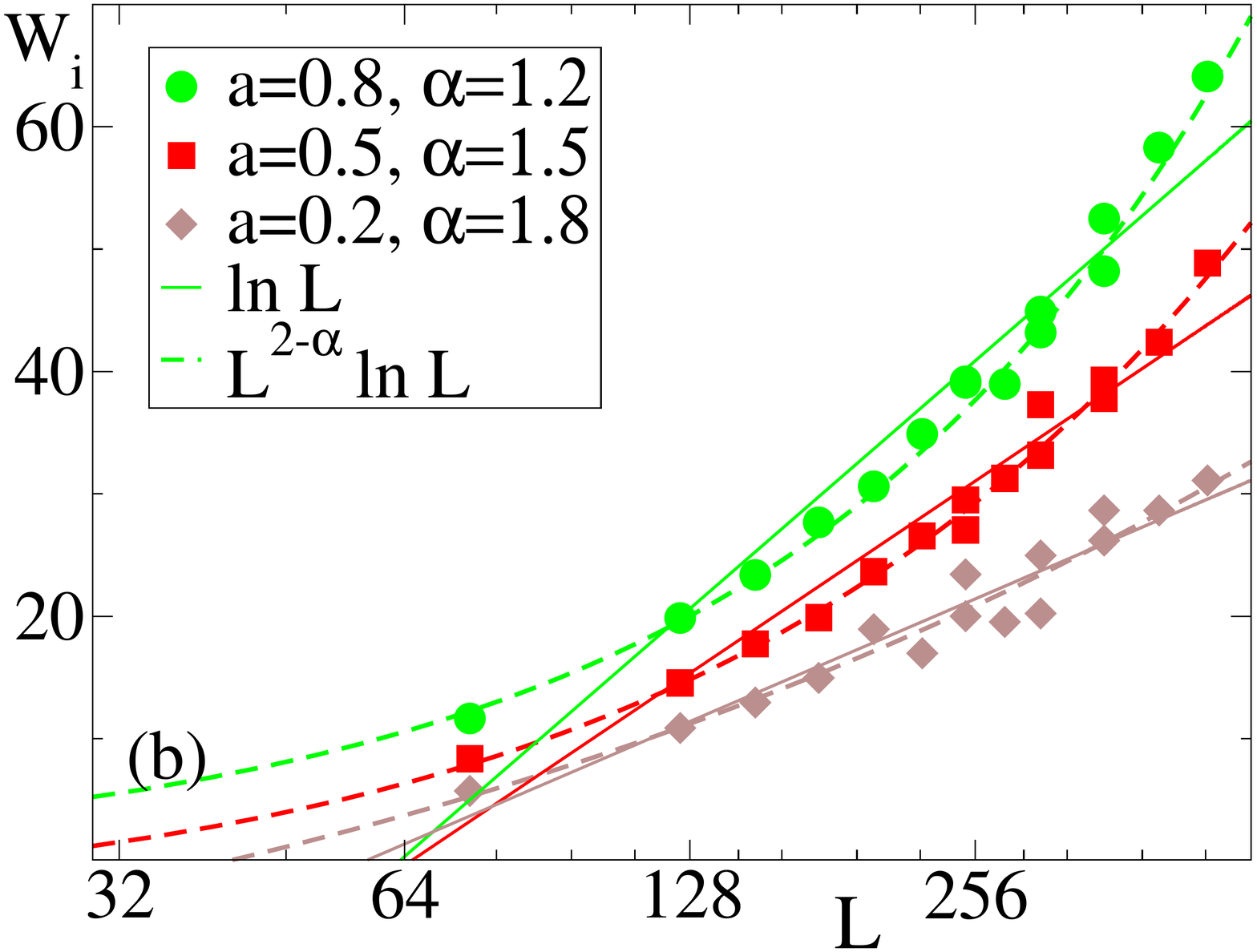}\\
\includegraphics[width=0.23\textwidth]{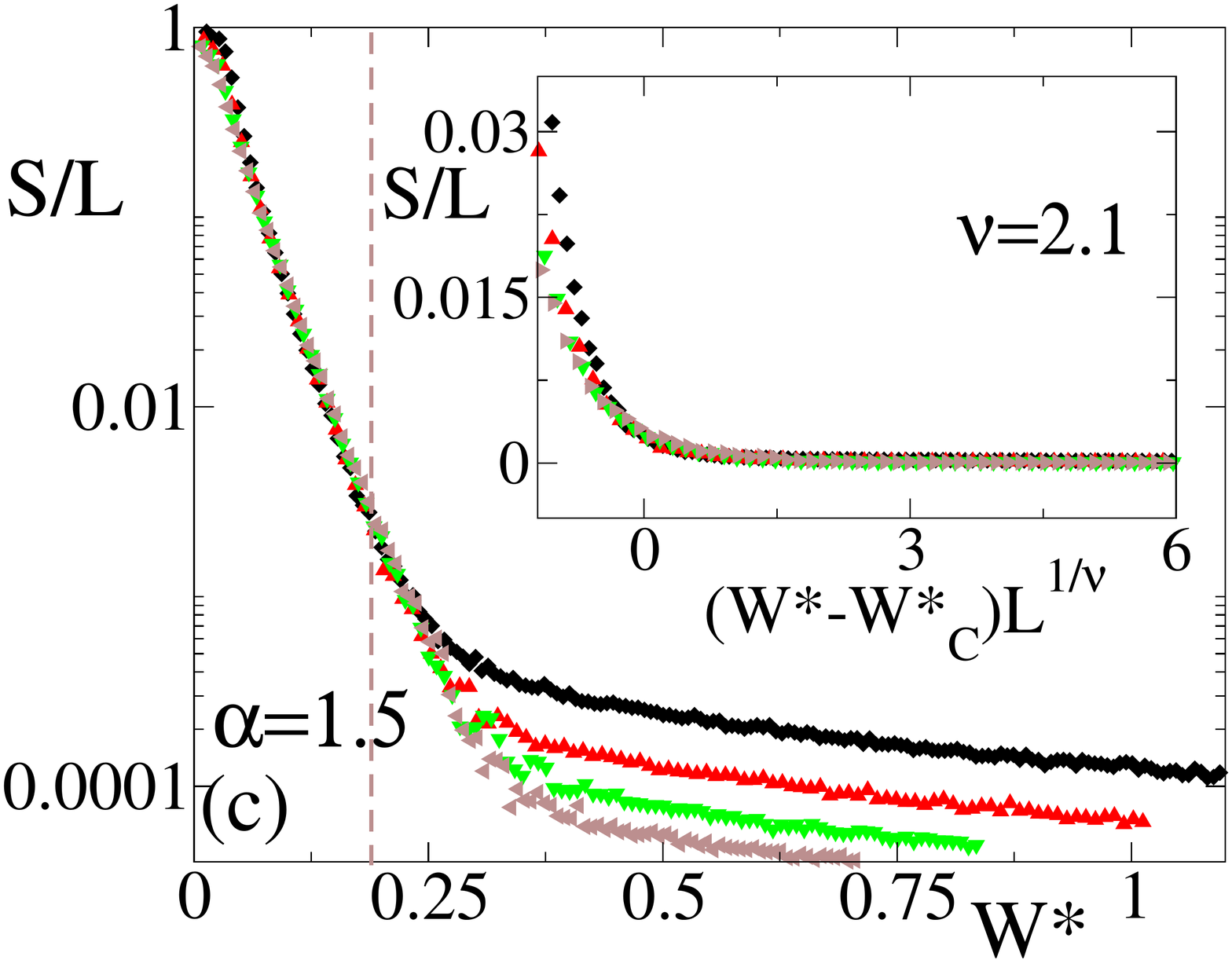}
\includegraphics[width=0.23\textwidth]{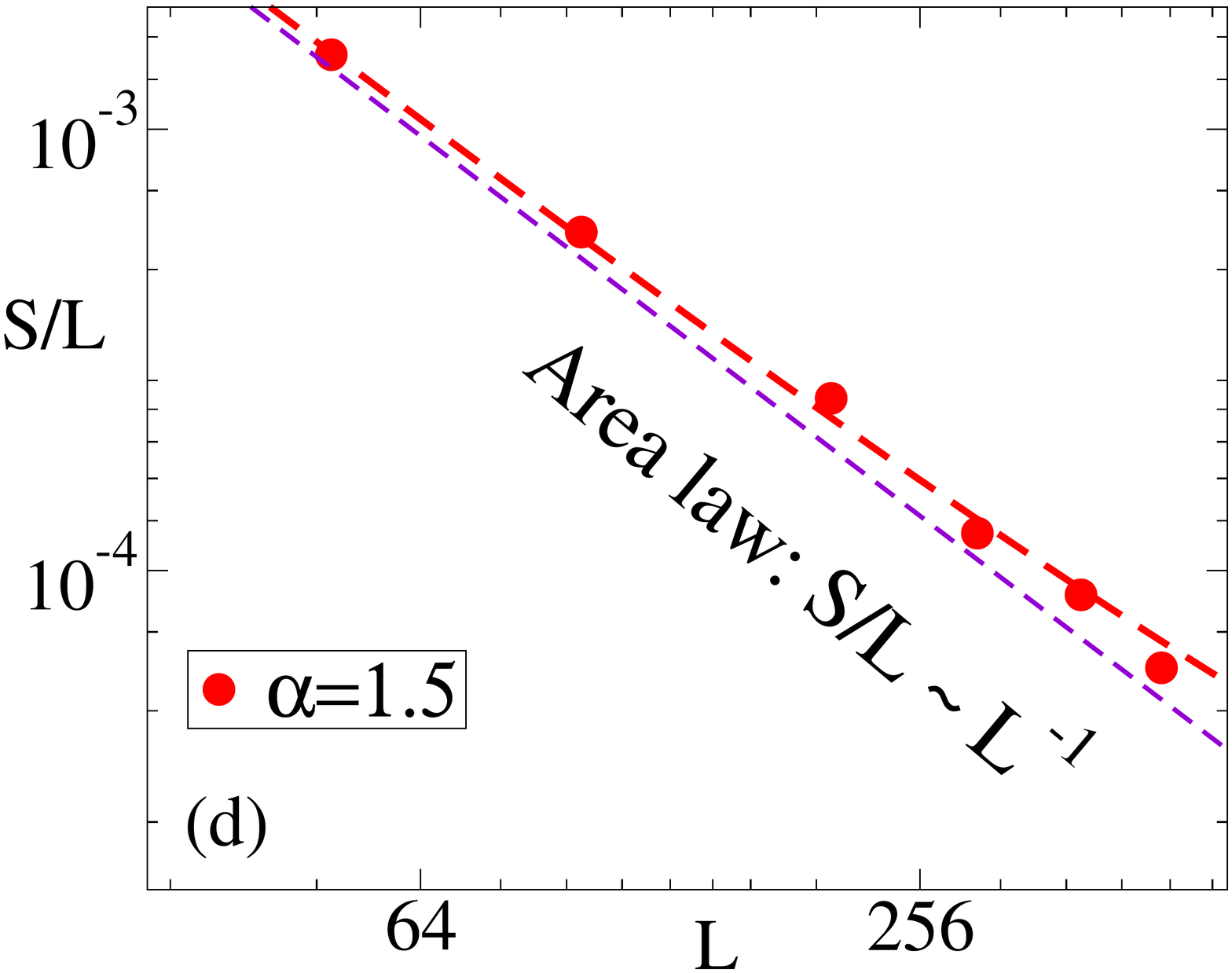}
\caption{(Color online) We here present the findings of EE putting
the microscopic detail of the Hamiltonian~\eqref{eq:micro_ham} in our RG scheme .
(a) EE density $S/L$ decreases from $1$
with increasing the  disorder strength $W$ for $L=100$, $200$, $300$, and $500$. Here $\alpha=1.2$. 
The intersection points of $S/L$ curve between different values of $L$,  shifts to higher values of $W$ as 
one increases system size. 
(b): Intersection points obtained from (a), depicted by solid red, green and brown points, scales with system size as $L^{2-\alpha}\ln L$
that is represented by the dashed lines. \textcolor{black}{The logarithmic 
fitting  ${\rm ln} L$ as depicted by solid lines are not are in good 
agreement with the numerical data points, specifically, for 
$\alpha=1.2$ and $1.5$.}
(c) We  again show the variation of $S/L$ with  rescaled disorder strength $W^{*}=W/L^{2-\alpha}\ln L$. 
We identify the transition point as $W^{*}_c=0.19$. In the inset, we show data collapse
$S/L=f((W^{*}-W^{*}_c)L^{1/\nu})$ for $\nu=2.25$ and for $W^{*}>W^{*}_{c}$. 
(d): We show a scaling of $S/L$ with $L$ in a log-log scale with $W=80.0$ as depicted by solid red points. 
The area law is guaranteed for $W^*=1.0$ 
in the localized phase as depicted by the dashed violet line.} 
\label{fig6}
\end{figure}

\subsection{Microscopic RG}
\label{s3ss2}

Even though, we believe our previous RSRG scheme in Sec.~ \ref{s3ss1} is able to capture the main essence of the long-range models. 
In this section,
we again do similar studies but now incorporate  the microscopic details of a  particular long-range Hamiltonian in the RSRG Scheme.
\textcolor{black}{Specially, using the microscopic RG scheme, 
the role of instability of MBL for $\alpha>2$ can probed more extensively even with finite size of system $L\sim O(10^2)$. }
The specific long-range microscopic model, we  use to modify the RG scheme  is described by the following Hamiltonian,
\begin{eqnarray}
 {H}=-\sum _{i,j\neq i}\frac{J_{i,j}}{(1+|i-j|^{2\alpha})^{1/2}}(\hat{c}^{\dag}_i\hat{c}^{}_{j}+\text{H.c.})+\sum _{i}\mu_i \hat{n}_i \nonumber \\
 \label{eq:micro_ham}
\end{eqnarray}
where $\hat{c}^{\dag}_i$ ($\hat{c}_{i}$) is the fermionic creation (annihilation) operator at site $i$,
$\hat{n}_i=\hat{c}^{\dag}_i\hat{c}_{i}$ is
the number operator, and $L$ is the size of the system. 
$J_{ij}$ and $\mu_i$ are uniform random number chosen from an interval $[-1,1]$ and $[-W,W]$ respectively. For $\alpha > 1$ the 
single particle states of this Hamiltonian  are algebraically localized. 
We first carry out ED calculation of this non-interacting Hamiltonian and obtain all single particle energies and 
eigenstates.
Given that a typical single-particle eigenstate with 
eigenenergy $\epsilon_i$ is of the form $\psi_i\sim1/|i-r_0|^{\alpha}$ , where $r_0\neq i$ is the localization center. We now initialize our 
RG scheme by defining $\Delta E_{ij}=|\epsilon_i -\epsilon_j|$ i.e., the
difference between the eigenenergies of the Hamiltonian (\ref{eq:micro_ham}). We consider
 $\Gamma_{ij}=1/|i_0-j_0|^{\alpha}$  where $i_0$, $j_0$ are the localization center correspond to 
the $i$-th  and $j$-th eigenstate of the Hamiltonian. 
{We note that the interacting version of this model has been studied where the MBL phase is
characterized by the algebraically decaying tails of an extensive number
of integrals of motion,  unlike the case of exponentially
localized SPSs \cite{tomasi.19}.}

In Fig.~\ref{fig6}(a), we show the variation entanglement density as 
function of $W$ for different values of $L$. Similar to the outcome from macroscopic scheme, 
the intersection point $W_i$ shifts to higher value with increasing $L$; we note that the window $\Delta W$ and the values of 
$W_i$ both acquire higher values compared to the earlier case. In order to search for the sharp transition 
point, we then try to estimate the proper scaling law of $W_i$ with $L$ 
in Fig.~\ref{fig6}(b) for $\alpha=1.2$, $1.5$ and $1.8$.  
 Interestingly, the microscopic input modifies the scaling function; it becomes 
  more rapid compared to the slow $\ln L$ scaling as shown in Eq.(\ref{eq:w_rescale}): 
  \be
  W^{*}= \frac{W}{L^{2-\alpha}\ln L}.
  \label{eq:micro_scale}
  \ee
This form of renormalization confirms the predicted scaling by  Mirlin etal. \cite{mirlin.18} following an ED
scheme in interacting  spin  model where hopping and interaction both considered to be long range.
{Even though, the microscopic input that we use here is from a  long-range non-interacting model (\ref{eq:micro_ham}), 
 but our RSRG scheme does not incorporate long-range interaction. Interestingly, we still 
manage to mimic the underlying physics of the long range model unanimously irrespective of the range of interaction  \cite{nag.19}}.
The scaling form (\ref{eq:micro_scale}) also matches well with the analytical prediction in the context of random regular graph \cite{gutman.16}
that we discussed in Sec.~(\ref{s3ss1}). Hence, microscopic detail in RG scheme helps in obtaining more accurate behavior 
for the observables. \textcolor{black}{We further check that logarithmic scaling 
$W_i \sim {\rm ln} L$ (as depicted by solid lines in Fig.~\ref{fig6} (b)) is not the accurate renormalization of $W_i$ for the microscopic 
RG scheme. Here, the algebraic scaling form prevails referring to the fact that 
finite size effect is relatively minimized compared to macroscopic RG scheme.}


\textcolor{black}{A very recent  ED study
shows that  long-range interaction is not able to influence the thermal-MBL
transition property of the system in a noticeable manner \cite{nag.19}. Contrastingly, MBL might  persist  in  the  presence
of  long-range interactions  though  long-range  hopping  with 
$1<\alpha<2$  delocalizes  the  system  partially,  while almost  all  
the  states are extended for $\alpha \le 1$.  On the other hand, our study of 
microscopic RSRG study  with short range diagonal interaction indicates the fact that critical disordering 
follow similar scaling with $L$  as observed in long-range interacting system. }


Similar to the Fig.~\ref{fig4}(a),
the EE density for different $L$ coincide with each other when $W^*<W^*_c=0.33$, after that they start
deviating from each other for $W^*\ge W^*_c$. For $W^*\gg W^*_c$,
the saturation values of EE density increases with decreasing $L$. Comparing Fig.~\ref{fig4}(a) and 
Fig.~\ref{fig6}(c), one can see that the tendency towards saturation is more once the RSRG scheme is embedded with the microscopic detail. 
Now we shall investigate the scaling form of $S/L=f((W^*-W^*_c)L^{1/\nu})$ from the data collapse analysis as shown in Fig.~\ref{fig6}(c). We
show here that with $\nu=2.25\pm 0.3$, one can obtain a nice data collapse for $W^*\ge W^*_c$. 
We also checked the critical properties  for $\alpha=2.5$ where 
we find exponent $\nu= 2.9\pm 0.27$.
Hence, these values of the critical exponent are well corroborated 
 to their counterpart obtained from macroscopic RSRG approach satisfying the Harris criterion\cite{harris1974effect}.
Therefore, microscopic RG reconfirms that the universality class for $\alpha<2$ is different than that of the for $\alpha>2$. 
{
Finally, in Fig.~\ref{fig6}(d), we show that the area law (violet dashed line) is recovered for localized phase in the the regime 
$W^* \gg W^*_c$.}

\begin{figure}
\includegraphics[width=0.5\textwidth]{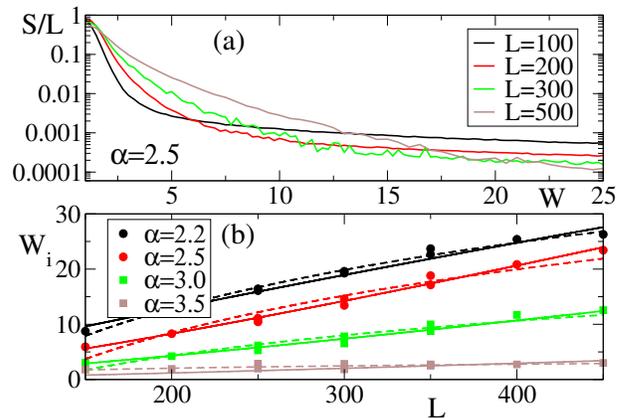}
\caption{\textcolor{black}{(Color online) The behavior of EE density as a function of 
disorder strength $W$ is shown for 
microscopic RSRG scheme for $\alpha=2.5$ in (a). It clearly suggests that 
there is finite delocalization-localization cross-over window in $W \{W_i\}$ for 
$\alpha=2.5$. We then try to estimate the behavior of $W_i$ with the 
system size $L$ considering two fitting function $m \ln L +n$ (dashed lines) and $p \exp(q
\sqrt{\ln L})$ (solid lines) as shown in (b).
One can observe that for $\alpha$ being well above $2$,  $p \exp(q
\sqrt{\ln L})$ yields apparently better fit compared to  $m \ln L +n$. 
We find $m=17.2,~16.5,
~9.1,~1.0$, and $n=-78.3,~-79.2,~-43.9,~-3.5$ for $\alpha=2.2,~2.5,~3.0,~3.5$, respectively.
We find $p=4.0\times10^{-4},~4.8
\times10^{-6},~2.5\times10^{-6},~0.9\times 10^{-7}$ and $q=4.5,~6.2,~6.1,~6.0$ for 
$\alpha=2.2,~2.5,~3.0,~3.5$, respectively. } }
\label{fig7}
\end{figure}

\textcolor{black}{
We now show the variation of $S/L$ with $W$ for $\alpha=2.5$ using the
microscopic RSRG approach in Fig.~ \ref{fig7}. Unlike the case of  the
macroscopic RSRG scheme, here we find that $W_c$ increases with $L$
even for $\alpha > 2$. However, the scaling of $W_c$ with $L$ becomes 
much slower compared to the one obtained before for $\alpha <
2$.  Figure.~\ref{fig7} shows the variation of $W_c$ with $L$ for
$\alpha=2.2$, $2.5$, $3.0$, and $3.5$. We find almost equally good agreement by
fitting our results with two functional forms 1) $\sim \ln L$ and 2)
$\sim \exp(\text{const}\sqrt{\ln L})$. The later one has been
predicted in Ref.~\cite{gopalakrishnan.19,mirlin.16} for $\alpha >2$ in presence of long-range
interactions.}

\section{Comparison between microscopic and macroscopic RSRG}
\label{comparison}
\textcolor{black}{
In this section, we compare  between  two RSRG schemes  which we have
used in this work. Even though, both RSRG approaches conclude the
critical disorder strength $W_c$
is an increasing function of $L$, 
the finite size scaling of $W_c$
are not very different in two approaches at least for $\alpha<2$.
While the microscopic RSRG predicts $W_c(L)\sim L^{2-\alpha}\ln L$ ,
the macroscopic RSRG predicts  $W_c(L)\sim L^{\eta} \ln L$ with $\eta \ll 2-\alpha $ for $\alpha <2$.
We believe that one of the main reason behind this discrepancy is due to
the  difference between the spacing distribution of the
on-site energies in two approaches. In the  microscopic RSRG scheme,
the initial on-site
energies have been taken from an uniform random distribution and
hence,  spacing distribution 
is Poissonian. On the other hand, in the microscopic RSRG approach,
the on-site energies are taken from
the eigenvalues of the microscopic Hamiltonian ~\eqref{eq:micro_ham}, the spacing
distribution of
eigen-energies of the Hamiltonian ~\eqref{eq:micro_ham} are not exactly
Poissonian~\cite{nosov.19}, they show the signature of level-repulsion at least for
$\alpha < 2$. Hence,  the microscopic RSRG scheme promotes
avalanche mechanism \cite{gopalakrishnan.19,roeck.17}. This is presumably the reason the scaling
$W_c(L)$ with $L$ is much faster in this scheme compared to macroscopic
RSRG scheme. }

\begin{figure}
\includegraphics[width=0.5\textwidth]{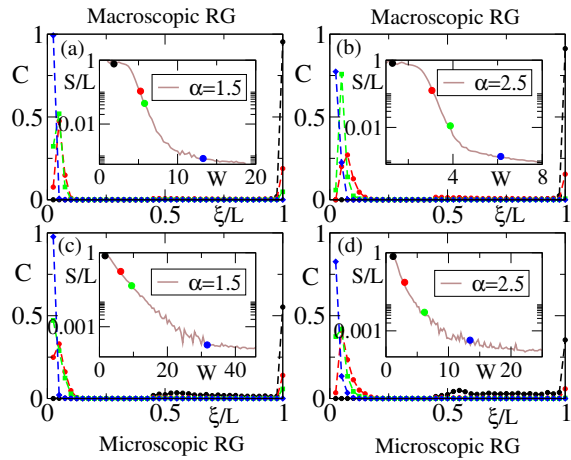}
\caption{ \textcolor{black}{
(Color online) We investigate the histogram of normalized MBS $C$ as a function of 
$\xi/L$ for microscopic RSRG with $\alpha=1.5$ (a), 
 $\alpha=2.5$ (b) and macroscopic RSRG with $\alpha=1.5$ (c), 
 $\alpha=2.5$ (d). We choose four representative points as 
 marked by black, red, green, blue solid circles in   
 inset where EE density is plotted with $W$ for $L=300$. 
 We show the variation of histogram count $C$  as a function of 
 $\xi/L$ for the above four representative points with specified 
 $W$. One can clearly observe from both of 
 the RSRG scheme that the distribution pattern changes from 
 uni-modal (when system remains  
 deep inside localized/ delocalized phase) to  
  bi-modal (when the system in the vicinity of delocalization-localization 
 cross-over window).
 For $\alpha=2.5$, disorder window of $W$ within which bi-modal distribution 
 appears, gets reduced compared to $\alpha=1.5$.
} }
\label{fig8}
\end{figure}

\textcolor{black}{
Now we investigate the distribution of the normalized MBS $\xi/L$ for
both RSRG schemes and  for $\alpha=1.5$ and $2.5$. We show our results
for $4$ different values of quenched disorder of strength $W$ in Fig.~\ref{fig8}.
One value of $W$ is chosen such that $W \ll W_c(L)$ (denoted with black
points in the Fig.~\ref{fig8}), for which the peak of the distribution is
at $\xi/L \simeq 1$  signifying complete delocalization.
Another value of $W$ is chosen such that $W \gg W_c(L)$ (denoted with
blue points in the Fig.~\ref{fig8}), for which the peak of the
distribution is at $\xi/L \ll 1$ and that signifies complete
localization. On the other hand, we also show the results for two
other values of $W$, which are taken from the vicinity of $W_c$ (denoted
with red and green points in the Fig.~\ref{fig8}).  The MBS distribution
shows bi-modal distribution having two peaks one at $\xi/L\simeq 1$
and another at $\xi/L \ll 1$.
The presence of cluster with MBS $\simeq L$ in this regime  might refer to the
instability of phase transition; the system   flows towards the
thermalizing phase due to its inbuilt avalanche mechanism. }

\textcolor{black}{
There is no significant qualitative differences 
between the distribution of MBS in two RG approaches. However,
the  inset plots of
$S/L$ vs $W$ suggest (insets of Fig.~\ref{fig8})
that within the microscopic RSRG approach the plateau region of
EE  for $W\ll W_c$ is absent ($S/L$ starts decreasing
as soon as $W$ is increased) in comparison to the macroscopic RSRG
scheme.  This leads to a  few quantitative 
differences while the distribution of
MBS is studied.
The height of the peak of MBS distribution  at $\xi/L\sim 1$ for
$W\ll W_c$  is much less ($C\sim 0.5$) for the  microscopic RSRG  compared to the
macroscopic RSRG scheme ($C\sim 1$).  
 In particular, for delocalized phase observed following microscopic RG, 
histogram of MBS $C$ gets distributed over a range $0.5\le \xi/L \le 1.0$ which is not observed for macroscopic RG.
On the other hand, the disorder window of cross-over region showing  the bi-modal distribution becomes shortened
for macroscopic RG compared to  microscopic RG. This enhancement in the cross-over window for microscopic RG again
refers to the fact that $W_c$-$L$ scaling can differ.
}

\begin{figure}
\includegraphics[width=0.5\textwidth]{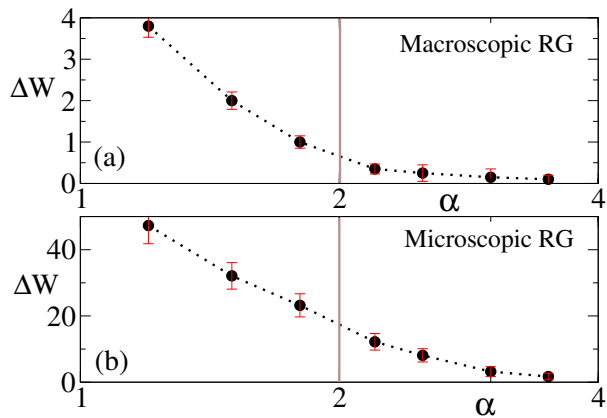}
\caption{ \textcolor{black}{
(Color online) We plot the crossover window in  disorder $\Delta W$,
with $\alpha$
obtained from macroscopic RG and microscopic RG in (a) and (b), respectively. 
We here use $L=100,\cdots,500$ data to estimate $\Delta W$. The distinct 
behavior is oberserved around $\alpha=2$ at least for (a).
} }
\label{fig9}
\end{figure}

\textcolor{black}
{{We  now show the behavior of the  cross-over window $\Delta W=W_i(L_{\text{max}})-W_i(L_{\text{min}})$
as a function of $\alpha$  for macroscopic RG and microscopic RSRG scheme
in Fig.~\ref{fig9} (a) and (b), respectively, where $L_{\text{max}}$ and $L_{\text{min}}$ are the 
largest and the smallest system sizes used for our numerical calculations. 
 It is evident from the plot  $\Delta W$ vs. $\alpha$ for macroscopic RSRG ,   
$\Delta W$ is significantly small for $\alpha> 2$ compared to $\alpha< 2$ . Hence, it is almost impossible  to  find 
finite size scaling of $W_i$ in this parameter regime. 
 On the other hand, within microscopic RSRG scheme $\Delta W$ remains relatively large even for $\alpha >2$, 
hence, it has been possible to study  the system size dependence of $W_c(L)$ in this parameter regime, 
which seems to satisfy the following scaling function:  $W_c(L) \sim \exp(\text{const}\sqrt{\ln L})$.}
In addition, we also note that area law of EE is satisfied for $\alpha>2$ for 
both of the cases without rescaling the disorder strength. Therefore, 
there are some qualitative changes occurring around $\alpha=2$ which need to be 
addressed more extensively in future.
}

\textcolor{black}{
We also like to emphasize that even within macroscopic RSRG approach $\Delta W$ for
$\alpha \ge 2$, is not strictly zero even for our choice of system sizes as shown in Fig.~\ref{fig9} (a). However, the variation of $W_i$  with  $L$   within macroscopic
RSRG approach is very small, when $L \in [50,500]$. Hence 
one does not even need to re-scale  $W$ with appropriate $L$ dependent scaling functions to see delocalization-MBL transition (as shown in Fig.~\ref{fig2} (b)).
This does not necessarily mean that MBL phase is stable in the thermodynamic limit. There is always a possibility that if we could manage to do our calculations for significant
large system sizes compared to what we have presented here,
we might obtain $\Delta W$ large enough to find system size dependence scaling function for $W_c$.
Strikingly, in the  microscopic RSRG approach even for $L \in [50,500]$, we find $\Delta W$  to be large enough to extract a $L$ dependent scaling function 
for $W_c$, which proves the instability of MBL phase in these models
in the thermodynamic limit. }


\section{Conclusion}
\label{s4}

In this work, we propose a new RSRG scheme to investigate thermal-MBL transition  in a one-dimensional  long-range models
 with hopping  $t\sim r^{-\alpha}$,
where SPSs are algebraically localized with localization exponent $\alpha >1$. 
Within this approach, \textcolor{black}{ in presence of nearest neighbour interaction},
we show that  indeed there is a crossover between delocalized and localized phase 
 as a function of  quenched disorder $W$ for $\alpha<2$.
In last few years, there have been several studies leading to conflicting claims
about the true nature of this transition \cite{burin.15,nag.19}.
Most of those studies involve ED that is
restricted within small system size.   Our RSRG approach allows  us to extend system size up to 
$L\simeq 500$, with which  we can investigate the finite size scaling of  transition points systematically.
Even though this  scaling   
appears to be dependent on RG scheme, the most realistic implementation of RG rules predicts
the scaling to be   $\sim L^{2-\alpha}\ln L$. This form  supports
the  prediction  of Ref.~\cite{mirlin.18}. We hence propose that one can still talk about thermal-MBL 
transition  in appropriate thermodynamic limit 
as function of  rescaled quenched disorder $W^{*}=W/L^{2-\alpha}\ln L$. Moreover,
the apparent deviation from the area law in the MBL phase is also  remarkably resolved upon 
considering $W^*$.
Most interestingly, with this non-trivial rescaling for $\alpha<2$,  we obtain different correlation length
exponents associated with the transition  which is qualitatively and
quantitatively different from a usual MBL transition observed in short
range system. On the contrary, the MBL transition for $\alpha>2$ requires no rescaling of $W$ and surprisingly,
it belongs to the same Anderson type universality class for the short
range systems.

\textcolor{black}{A recent study claims the range of interactions does not influence the thermal-MBL transitions 
as long as long-range hopping is present in the systems  ~\cite{nag.19}. The finite size scaling of the critical disorder strength obtained from
our calculations matches with analytical prediction for long-range interacting models  ~\cite{mirlin.18, gopalakrishnan.19}.
Our 
findings qualitatively supplement ED results along with the possibility that 
$\alpha=2$ can be a special point. 
However, for long-range hopping system without 
diagonal interaction, the  MBL  transition might 
occur for $a>3/2$ \cite{maksymov20};  with diagonal 
interaction the crossover point  
$a=3/2$ is not found to be conclusive \cite{yao.14}. 
Combining all these above discussions, we can infer that the location of the exact 
crossover point for short range diagonal interaction (long-range hopping) 
is a subject of further study.}

\textcolor{black}{One can note that the apparent dissimilarity in the scaling of critical disorder 
for macroscopic and microscopic  RSRG, might be caused by the 
average level spacing statistics of the initial input energies of
these model e.g., macroscopic  RSRG supports Poissonian distribution while {for microscopic RSRG 
it deviates towards Wigner-Dyson distribution. This 
 promotes the avalanche mechanism \cite{gopalakrishnan.19,roeck.17} in the system, and destabilizes the MBL phase faster.}}
\textcolor{black}{In this context of stability of MBL phase for short range system, it is also important to 
mention the recent studies. These   
 suggest that in order to observe a MBL transition, one needs to consider 
 thermodynamically large systems supporting Thouless time scale comparable to Heisenberg 
 time scale \cite{lev.2019,abanin19,sierant19,Panda_2020}. We believe that it would be interesting to investigate these time scales for  
  long-range systems as well.}

The statistics of many-body energy levels of long-range systems is experimentally 
investigated in superconducting circuit \cite{Roushan.17} and
trapped ion \cite{haffner.05, zhang.17,
bernien.17}. On the other hand, long range hopping is also
realized in laboratory \cite{childress2006, Ni.08}. We therefore believe that
our findings can be experimentally testable in near future.
One natural extension to our work would be to analyze the effect of long
range interaction and probe the MBL transition
\textcolor{black}{which has already been investigated using self-consistent theory and ED \cite{sierant.19,roy2019self}}.
An sub-extensive law of EE in the localized phase is clearly
observed for non-interacting system \cite{modak.19}; then the question becomes in presence of interaction is this
law  suppressed and EE tends toward the area law. Hence, a possible future direction would be to
critically analyze the scaling of EE in a thermodynamically  large
system with various other RSRG scheme incorporating appropriate microscopic detail.
On the other hand, the existence of Floquet time crystal
in this long range model  can be another field of research.


\section{Acknowledgement} 
We would like to thank Andrew C. Potter for initial discussion regarding the RSRG algorithm. We also thank Anirban Roy for 
helping us with some relevant python packages during our numerical calculation.
 
\bibliography{reference}

\end{document}